\documentclass[preprint]{revtex4-1}
\usepackage{mathrsfs}
\usepackage[latin1]{inputenc}
\usepackage{amsmath,amssymb}
\usepackage{graphicx}
\usepackage{subcaption}
\usepackage{float}
\usepackage{mwe}

\newcommand{\ben}{\begin{equation}}
\newcommand{\een}{\end{equation}}
\newcommand{\bea}{\begin{eqnarray}}
\newcommand{\eea}{\end{eqnarray}}

\def\ket#1{\vert#1\rangle}

\def\sss{\scriptscriptstyle\rm}

\def\1s{_{1,\sss S}}
\def\2s{_{2,\sss S}}

\def\br{{\bf r}}

\def\bA{{\bf A}}

\def\dulr{{\underline{\underline{\bf r}}}}
\def\dulR{{\underline{\underline{\bf R}}}}
\def\dulq{{\underline{\underline{q}}}}
\def\dulQ{{\underline{\underline{Q}}}}

\usepackage{ulem}

\begin{document}
\title{Exact Factorization Adventures: A Promising Approach for Non-bound States}
\author{Evaristo Villaseco Arribas$^1$, Federica Agostini$^2$,}
\author{Neepa T. Maitra$^1$}
\email{neepa.maitra@rutgers.edu}
\affiliation{$^1$Department of Physics, Rutgers University\\
Newark, New Jersey 07102, USA\\
$^2$Universit\'e Paris-Saclay, CNRS, Institut de Chimie Physique UMR8000\\
 91405 Orsay, France}

\begin{abstract}
Modeling the dynamics of non-bound states in molecules requires an accurate description of how electronic motion affects nuclear motion and vice-versa. The exact factorization (XF) approach offers a unique perspective, in that it provides potentials that act on the nuclear subsystem or electronic subsystem, which contain the effects of the coupling to the other subsystem in an exact way. We briefly review the various applications of the XF idea in different realms, and how features of these potentials aid in the interpretation of two different laser-driven dissociation mechanisms. We present a detailed study of the different ways the coupling terms in recently-developed XF-based mixed quantum-classical approximations are evaluated, where either truly coupled trajectories, or auxiliary trajectories that mimic the coupling are used, and discuss their effect in both a surface-hopping framework as well as the rigorously-derived coupled-trajectory mixed quantum-classical approach.
 \end{abstract}

\maketitle

\section{Introduction}
A challenge in the theoretical simulation of molecules driven out of their ground-state by a strong field is how to account for, and interpret, the coupling between the electrons and ions. If a molecule was only gently perturbed at frequencies below electronic excitations, the Born-Oppenheimer (BO) picture can still be useful: the disparity in the time-scales of nuclear and electronic motion allows for the nuclei to evolve on a single BO surface for much of the time, and regions where this  breaks down tend to be somewhat localized in space and involving the interaction of only a small number of BO states. In those regions, a fully quantum treatment of the molecule based on the BO picture would involve propagating nuclear wavefunctions projected on the different surfaces, including coupling terms between them. 
If the nuclear degrees of freedom were to be treated classically, the effective force driving their motion would deviate from that of a single surface, or even a mean surface: different parts of the underlying nuclear wavepacket would be associated with different electronic surfaces, even within the same spatial region. The problems with capturing this electron-nuclear correlation in such mixed quantum-classical methods are well-known~\cite{CB18,AC19}. But with strong fields, or with high-frequency fields, the problem is even more challenging: the nuclear wavepacket straddles many BO states, a continuum beyond the ionization threshold. Field-dressed surfaces such as quasi-static~\cite{TIW96,TIW98,BGSSRMG12} or Floquet~\cite{HSS01,FHS16,Agostini_JCP2021_2} can be instructive in this regard; although both serve equally well as a basis for the nuclear dynamics, as does the BO basis, approximations inherent in mixed quantum-classical approaches may perform better in one or the other,  for physical reasons or computational reasons, but for general field parameters, the jury is out.  Further, the observables of interest depend on the application: for example, we may want to  correlate electron ionization and nuclear dissociation channels to get a full understanding of the dynamics~\cite{Vrakking_NP2014}, or, we may be primarily interested in the behavior of the ionizing electron itself, with a time-dependent potential driving the electron helping to guide our understanding. As intuitive and instrumental the BO picture has been, it does not provide such a potential, nor is it convenient in strong fields when many surfaces get involved. 

An alternative, but equally fundamental, picture is provided by the exact factorization (XF) approach~\cite{Hunter75,Hunter_IJQC1975_2,Hunter_IJQC1980,H81,Hunter_IJQC1982,GG14,AMG10,AMG12,AG21}. XF re-casts the dynamics of interacting quantum subsystems in a way that provides unique potentials that drive the motion of each subsystem and account for complete coupling to others. Applied to the electron-nuclear case, the exact molecular wavefunction $\Psi(\dulr, \dulR,t)$ is factored into a single correlated product of a marginal factor and a conditional factor. Choosing the marginal to be a function of $\dulR$, one obtains a time-dependent Schr\"odinger equation (TDSE) for the nuclear wavefunction, in which the potentials contain the complete effect of coupling to the electrons~\cite{AMG10}. Instead, choosing the marginal to be a function of $\dulr$ only, a TDSE is obtained for the electronic wavefunction, in which the potentials contain the complete effect of coupling to the nuclei~\cite{SAMYG14}. These potentials drive the nuclear and electronic subsystem respectively, and have been instructive in interpreting correlated electron-ion dynamics, including for dissociation~\cite{AMG10} and ionization processes~\cite{KAM15} in strong fields. To be a practical approach, approximations must be made, and recently two mixed quantum-classical (MQC) approaches have been derived, in which a key ingredient in the electron-nuclear coupling terms is the so-called nuclear quantum momentum. The quantum momentum measures the spatial variation of the nuclear density during non-adiabatic dynamics, and couples strongly to the electronic evolution giving large corrections to mean-field approximations. Different schemes have proposed different ways to calculate this term, using either the support of coupled trajectories or of auxiliary trajectories to track the delocalization of the nuclear density, but until now a detailed comparison has not been made. Here we compare the use of auxiliary trajectories within an XF-based surface-hopping framework (SHXF)~\cite{HLM18,PyUNIxMD} to using coupled trajectories within the surface-hopping procedure (CTSH)~\cite{PA21} as well as within the original coupled-trajectory mixed quantum-classical approach (CTMQC)~\cite{MAG15,AMAG16,MATG17} that was derived directly from the XF. We further study the impact of a modified definition of the quantum momentum in the coupled-trajectories calculations CTSH and CTMQC, that ensures the physical condition of zero population transfer in regions of no non-adiabatic coupling.

The paper is organized as follows. In Sec.~\ref{sec:XF}, we review the formalism of the XF approach, and briefly meander down various avenues in which the XF idea has been extended and explored. Sec.~\ref{sec:diss_ion} demonstrates its usefulness for the dynamics of non-bound systems, briefly reviewing the results of Refs.~\cite{AMG10,AMG12} for dissociation and Refs.~\cite{KAM15,KARM17} for ionization, in a one-dimensional model of H$_2^+$. The most practical impact of the XF approach is the development of rigorous MQC methods, and in Sec.~\ref{sec:MQC} we outline the CTMQC, CTSH, and SHXF schemes, before delving into the different ways the quantum momentum is computed in different implementations of these schemes in Sec.~\ref{sec:example}. We offer some conclusions in Sec.~\ref{sec:concs}. 

\section{The XF Approach in a Nutshell}
\label{sec:XF}
Consider the time evolution of a system of coupled quantum subsystems, possibly subject to some external fields:
\ben
{\hat H}\Psi(\dulq,\dulQ,\ldots,t) = i \partial_t\Psi(\dulq,\dulQ,\ldots,t)
\een
(Atomic units, $e^2 = \hbar = m_e = 1$, are used throughout this article unless otherwise stated). 
 These could be different types of particles, for example with $\dulq$ representing electronic coordinates  and $\dulQ$ nuclear coordinates, and perhaps another set of coordinates for photons.  Or, for example, we could have a system of the same type of particle, say electrons, but with $\dulq$ being occupation numbers of some set of orbitals, {\it e.g.} weakly-correlated and $\dulQ$ being occupation numbers of strongly correlated orbitals.  The problem is quite general, and in the XF the full wavefunction is factored into a marginal term and a conditional term~\cite{Hunter75, GG14, AMG10,AMG12}:
  \ben
 \Psi(\dulq,\dulQ,t) = \chi(\dulQ,t)\Phi_\dulQ(\dulq,t), \; {\rm where} \;\;\langle\Phi_\dulQ(t)\ket{\Phi_\dulQ(t)}_\dulq = 1 \;{\rm for \; all} \;t, \dulQ
 \label{eq:XF}
 \een
 Here $\langle ...\ket{...}_\dulq$ represents an inner product over all $\dulq$ variables only. 
The factorization Eq.~(\ref{eq:XF}) is exact and unique up to a $\dulQ$- and $t$-dependent phase, where $e^{i\theta(\dulQ,t)}$ multiplies $\chi$ while $e^{-i\theta(\dulQ,t)}$ multiplies $\Phi_\dulQ$, yielding the same product. 
For Hamiltonians that have the form
\ben
{\hat H} = -\frac{1}{2}\sum_I{\frac{1}{M_I}}\nabla^2_{Q_I}-\frac{1}{2}\sum_i{\frac{1}{m_i}}\nabla^2_{q_i} + V(\dulq,\dulQ,t)
\label{eqn: general H}
\een
where  $V(\dulq,\dulQ,t)$ is a purely multiplicative operator in $\dulq, \dulQ$, applying the Dirac-Frenkel variational principle shows  that the marginal factor $\chi(\dulQ,t)$ satisfies a TDSE with a scalar and vector potential that depend on $\Phi_\dulQ(\dulq,t)$ while the conditional factor satisfies a more complicated equation, with coupling terms dependent on $\chi(\dulQ,t)$~\cite{AMG10,AMG12,AMG13}. In Eq.~(\ref{eqn: general H}) we used the symbols $M_I$ and $m_i$ to indicate the masses -- in atomic units -- of the two sets of particles with positions $\dulQ$ and $\dulq$, respectively. Specifically, for the electron-nuclear problem, with $\dulQ = \dulR$ the nuclear coordinates, and $\dulq = \dulr$ the electronic coordinates, and the partial normalization condition $\int d\dulr\vert\Phi_{\dulR}(\dulr,t)\vert^2=1$, the wavefunctions $\Phi_\dulR(\dulr,t)$ and $\chi(\dulR,t)$ satisfy:
\bea
  \label{eq:exact_el_td}       
&&\Bigl(\hat{H}_{el}(\dulr,\dulR,t)-\epsilon(\dulR,t)\Bigr)\Phi_\dulR(\dulr,t)=i\partial_t \Phi_{\dulR}(\dulr,t) \\
  \label{eq:exact_n_td}                           
&&\Bigl(\sum_{\nu=1}^{N_n}\frac{1}{2M_\nu}(-i\nabla_\nu+{\bf A}_\nu(\dulR,t))^2 +\hat{V}_n^{ext}(\dulR,t) + \epsilon(\dulR,t)\Bigr)\chi(\dulR,t)=i\partial_t \chi(\dulR,t)           
\eea
with the electronic Hamiltonian $\hat H_{el}$ defined as
\bea
\label{eq:e_ham_td}
\hat{H}_{el}(\dulr,\dulR,t) = \hat{H}_{BO}(\dulr,\dulR)+ \hat{U}_{en}[\Phi_\dulR,\chi]+\hat V^{ext}_e(\dulr,t)
\eea
and with electron-nuclear coupling term
\begin{align}
\hat{U}_{en}[\Phi_\dulR,\chi]=\sum_{\nu=1}^{N_n}\frac{1}{M_\nu}\left[\frac{(-i\nabla_\nu-{\bf A}_\nu(\dulR,t))^2}{2} + \left(\frac{-i\nabla_\nu \chi}{\chi}+{\bf A}_\nu(\dulR,t)\right)\left(-i\nabla_\nu-{\bf A}_\nu(\dulR,t)\right)\right]
\label{eq:Uen}
\end{align}
Nuclear masses in atomic units are indicated with the symbol $M_\nu$, with $\nu$ labelling the nuclei. Here $\hat{H}_{BO}$ is the usual BO Hamiltonian (the sum of the electron kinetic energy, electron-electron, nuclear-nuclear, and electron-nuclear interaction operators). The  $\hat V^{ext}_n(\dulR,t)$ or $\hat V^{ext}_e(\dulr,t)$ potential is any externally applied potential, such as a laser field, acting on the nuclei or the electrons, respectively. We denote the time-dependent scalar potential appearing in Eqs.~(\ref{eq:exact_el_td}) and~(\ref{eq:exact_n_td})
\ben
  \label{eq:exact_eps_td}
\epsilon(\dulR,t) = \left\langle\Phi_\dulR(t) \left\vert\hat{H}_{el}(\dulR,t) - i \partial_t\right\vert \Phi_\dulR(t)\right\rangle_{\dulr}
\een
as the time-dependent potential energy surface (TDPES). Together with the time-dependent vector potential appearing in Eqs.~(\ref{eq:exact_n_td}) and~(\ref{eq:Uen})
\ben{\bf A}_\nu(\dulR,t)=\left\langle\Phi_\dulR(t)\right\vert\left.-i\nabla_\nu\Phi_\dulR(t)\right\rangle_{\dulr}\;,
\label{eq:exact_A_td}
\een
and the electron-nuclear coupling term $\hat{U}_{en}$, these three terms 
embody the exact electron-nuclear coupling. 
Refs.~\cite{AMG10,AMG12} showed Eqs.~(\ref{eq:exact_el_td}-\ref{eq:exact_A_td}) are form-invariant under
a gauge-like transformation that multiplies $\Phi_\dulR$ by an
$\dulR,t$-dependent phase $e^{i\theta(\dulR,t)}$, and $\chi$ by its inverse; the potentials transform correspondingly as $\bA \to \bA + \nabla\theta(\dulR,t)$ and $\epsilon(\dulR,t) \to \epsilon(\dulR,t) + \partial_t\theta(\dulR,t)$. Further, it was shown that we can identify $\chi(\dulR,t)$ as the nuclear wavefunction, in the sense that its modulus-square gives the $N_n$-body nuclear density, and the exact $N_n$-body nuclear current-density can be extracted in the usual way from its phase together with the vector potential~\cite{AMG10}. 

These equations emphasize the distinction from the far simpler BO approximation, that also involves a single correlated product, but with a time-independent electronic wavefunction and much simpler potential terms. The TDSE-form of the nuclear equation means that the potentials appearing in it can be directly interpreted as driving the nuclear motion, fully incorporating the effect of non-adiabatic and time-dependent coupling to the electronic system. 

Notice that the potentials appearing in the nuclear equation, $\epsilon(\dulR,t)$ and $\bA_\nu(\dulR,t)$,  involve the electronic wavefunction $\Phi_\dulR(\dulr,t)$, while the electron-nuclear coupling term  appearing in the electronic equation, $\hat{U}_{en}$, involves the nuclear wavefunction $\chi$: thus the evolution of each factor is intimately correlated with the other.


\subsection{XF Adventures in Different Realms}
\label{sec:adventures}
Although originally presented for coupled electron-nuclear dynamics, the XF approach applies quite generally to interacting quantum subsystems. It has been explored and extended along myriad different paths over the past decade. We summarize some of these here. 

{\it Exact potentials driving the nuclear motion} The first applications~\cite{AMG10,AMG12} showed the usefulness of the XF approach in providing potentials that directly correlate with nuclear motion in laser-driven dynamics, aiding in the interpretation of the physical processes. We defer further discussion of this to Sec.~\ref{sec:diss_ion}. The exact surfaces have been instructive for analysis of basis-dependence of MQC surface-hopping schemes for laser-driven bond-softening and bond-hardening processes~\cite{FHGS17}, and control of laser-induced electron localization~\cite{SAMG15}.  
For field-free dynamics following a photo-excitation,  the exact potential was found to track the BO surface initially occupied, develop a diabatic-like  character as the wavepacket passed through a non-adiabatic coupling region and then form steps that bridged the different BO surfaces associated to the local nuclear wavepacket~\cite{AASG13,AASG13b,AASMMG15,Agostini_JPCL2017}. The steps appeared in a gauge-dependent term which, when gauge-transformed to a vector potential, could be interpreted as a momentum-rescaling. Interestingly, in situations where nuclear wavepackets associated with different electronic states meet, the exact TDPES develops features such that independent classical nuclear trajectories evolving on it mimic the true nuclear distribution of interfering nuclear wavepackets \cite{CAG16}. In all these studies,  the exact potentials were obtained from an inversion of the numerical solution of  the full molecular TDSE. 

{\it Mixed quantum-classical and quantum trajectory methods} The most practical impact that the XF has had so far has been in the development of rigorous MQC methods \cite{AAG14,AAG14b,MAG15,AMAG16,MATG17}. We defer a discussion of this to Sec.~\ref{sec:MQC}.
While a purely quantum-classical scheme is not able to capture correctly nuclear quantum effects, such as tunnelling through a potential barrier of light particles~\cite{Heller_JCP1981, Burghardt_JPCA2007, Vuilleumier_PCCP2013, Rossi_FD2020, Manolopoulos_FD2020, Blumberger_FD2020, Franco_JCP2017, Tully_JCP2012} or scattering processes at low temperatures~\cite{DUPUY2022139241}, the use of \textit{quantum trajectories} within an XF framework~\cite{Suzuki_PRA2016, Ciccotti_JPCA2020, Ciccotti_EPJB2018} holds the promise to overcome this limitation while maintaining a general non-adiabatic viewpoint. Proof-of-principle studies in this direction have been proposed employing a Bohmian perspective~\cite{Lopreore1999, Wyatt1999, Wyatt2001, wyattbook} on nuclear dynamics with time-dependent XF potentials, and developments are currently ongoing especially aiming to address well-known numerical instabilities~\cite{Garashchuk_JCTC2019, Garashchuk_JPCC2010, Bittner_JCP2000, Wyatt_CPL2002, Kendrick_JCP2003, Wyatt_JCP2003} inherent a quantum-trajectory method. Similarly, using a hydrodynamic description of the XF equations, a trajectory-based representation for the electrons had been introduced in Ref.~\cite{Schild_arXiv2021} coupled to quantum nuclei.

{\it Inclusion of spin-orbit coupling} Non-adiabatic processes mediated by spin-orbit coupling (SOC), {\it i.e.} intersystem crossings,  play a key role in unraveling many excited-state processes such as photo-induced dynamics in organometallic complexes~\cite{Rothlisberger_CP2011, Gonzalez_CEJ2020, Gonzalez_JCTC2017, Boggio-Pasqua_M2017, Sato_CPL2012, Daniel_CTC2014} and even collision reactions involving systems of light elements~\cite{Schatz_JPCA2008, Casavecchia_PNAS2012, Agostini_JPCA2021}.  Refs.~\cite{TMRLA20,TSRLA20} 
extended the XF approach to derive a MQC approach that addresses both spin-allowed electronic transitions mediated by non-adiabatic coupling (NAC) as well as spin-forbidden transitions mediated by the SOC in the same theoretical construction. 
 
{\it Geometric phase} The concept of a geometric phase arises naturally in XF and can be defined as the circulation integral of the vector potential along a closed path, both in the stationary formulation of XF~\cite{MAKG14, RTG16, Requist_PRA2017} and in its time-dependent version~\cite{CA17, AC18, Agostini_JPCA2022, HAGE17}. It is worth underlining, though, that a non-vanishing geometric phase (not topological, as it is not quantized) in the framework of XF does not require invoking the adiabatic separation of electronic and nuclear motion in a molecule, thus it is a more general concept that does not rely on the particular representation (adiabatic vs. diabatic, for instance) used to describe the electronic subsystem.

{\it Density functionalization} The XF has been used to extend density functional theory (DFT) to non-adiabatic situations, where the conditional electron density $n_\dulR(\dulr) = N_e\int \vert\Phi_\dulR(\dulr, \dulr_2...\dulr_N)\vert^2 d^3r_2...d^3r_N$ replaces the $N_e$-body wavefunction as basic variable~\cite{RG16} of the theory. The full nuclear wavefunction $\chi(\dulR)$ is retained, and the  exchange-correlation functional of the usual adiabatic DFT is generalized to include nonadiabatic contributions arising from the $\hat{U}_{en}$ coupling term; it has functional dependence on the conditional electronic density and paramagnetic current-density, nuclear wavefunction, vector potential, and the quantum geometric tensor.  Ref.~\cite{LRG18} proposed a local conditional density approximation, and  applied the XF-based  DFT to study dissociation in a Hubbard-model of the LiF molecule, showing that non-adiabatic electron-nuclear coupling results in a 0.5 Bohr elongation of the bond-length at which electron-transfer is onset.

{\it Exact electron factorization} The XF formalism can be applied within a purely electronic system, where the marginal factor is a function of one electronic coordinate $\br_1$, with the conditional factor a function of the rest, conditionally dependent on $\br_1$~\cite{SG17}. From the partial normalization condition and the antisymmetry of the full electronic wavefunction, it follows that the marginal density gives the exact one-body electronic density and current-density.  The exact XF potentials for an electron ionized by a field have been studied~\cite{SG17,KS20,Schild_PRR2020}, and approximated, and  they
provide a new perspective for one-electron approaches commonly used in strong-field physics, such as the single active electron approximation. 

{\it Electronic embedding: strong correlation} 
The XF idea can also be applied to an electronic wavefunction in Fock space~\cite{GZR18,LM20,RG21}, where some orbitals ({\it e.g.} the more strongly correlated ones) are selected to be those of the marginal (the ``fragment") while all others are conditionally dependent on the occupation numbers of the marginal. 
XF then yields an embedding Hamiltonian on the fragment, which is  approximated by solving the full system at a low-level; the embedded fragment is then solved with a high-level method. This has been demonstrated to be accurate for ground-state energies over a full range of weak to strong correlation in Hubbard model systems~\cite{LM20}. 
A generalized Kohn-Sham approach has also been used to find the conditional factor~\cite{RG21}, using an orbital-dependent functional approximation, which accurately captured the topological phase diagram of a multiband Hubbard model. 

{\it Reverse factorization} 
There is nothing in the factorization presented above that requires us to choose the marginal as the nuclear coordinate, and the conditional as the electronic. It would be formally equally possible to choose the marginal factor to be the wavefunction for the electronic coordinate, and then the nuclear wavefunction is conditionally dependent on the electronic one. Although less intuitive, this is particularly useful for problems where we are primarily interested in electronic dynamics, since it yields a TDSE for the electrons, in which the potentials contain the effect of coupling to the nuclei. This was first studied in Ref.~\cite{SAMYG14} with the example of laser-control of electron localization: the time-delay between the excitation of a nuclear wavepacket from the ground to the first excited state of H$_2^+$ and an applied infrared pulse controls the localization asymmetry, {\it i.e.} the probability the electron ends up on the ``left" or ``right" nucleus.
We will return to this briefly in Sec.~\ref{sec:diss_ion}~\cite{SAMYG14,KAM15,KARM17}. 

{\it Mathematical analysis} Formally, the XF expression of the time-dependent molecular wavefunction closely resembles the BO approximation. However, in the latter case, the electronic conditional term is constrained to be an eigenstate of the BO Hamiltonian $\hat H_{BO}(\dulr,\dulR)$ all along the dynamics and, thus, it does not depend on time. Ref.~\cite{EA16} showed that the BO approximation can be recovered from XF as a limiting case for the electron-nuclear mass ratio $\mu$ tending to zero. 
Not only is this procedure an elegant scheme to connect XF to the well-known BO approximation, but it has been employed as well to develop algorithms that solve the coupled electronic~(\ref{eq:exact_el_td}) and nuclear~(\ref{eq:exact_n_td}) equations of XF. More specifically, the expansion of those equations in powers of $\mu$ helped in identifying leading non-adiabatic contributions in the electronic equation~(\ref{eq:exact_el_td}) to be retained in the approximate CTMQC algorithm (see Sec.~\ref{sec:MQC}), as well as in suggesting strategies for a perturbative treatment of non-adiabatic effects~\cite{Schild_JPCA2016, Scherrer_PRX2017, Scherrer_JCP2015, Gross_PRB2019} (discussed below).

It is of fundamental interest whether the set of equations, Eqs.~(\ref{eq:exact_el_td}-\ref{eq:exact_A_td}), are numerically stable to propagate, but it is also of practical interest, since in making approximations, we would like to know which terms might need care in developing methods that converge stably. In fact the mathematical form of the equations is unprecedented, and the usual numerical methods for TDSE's fail when applied in a straightforward way to the exact equations~\cite{GLM19}; a deeper mathematical analysis can be found in Ref.~\cite{Lorin21}.

{\it Polaritonic chemistry} The application of XF to systems of coupled electrons, nuclei, and photons~\cite{HARM18}, has provided new perspectives in the burgeoning field of polaritonic chemistry~\cite{YXS22}. Confining a molecule to an optical or plasmonic cavity enhances the light-matter coupling strength that scales as the square-root of the volume, and distorts potential energy surfaces according to how close the energy gaps at a given nuclear configuration are to the resonant frequencies of the cavity, thus altering reaction pathways. Choosing the marginal factor for the nuclear coordinate provides the exact TDPES that drives the nuclear motion, directly correlating with the nuclear dynamics in contrast to the polaritonic surfaces~\cite{GGF15} which are generalizations of the BO surfaces; Ref.~\cite{LHM19} showed how its features result in the suppression of proton-coupled electron-transfer in a model system, unlike the polaritonic surfaces. To properly model the complex systems used in the experiments, approximations such as MQC would be required, and running classical dynamics on the exact TDPES~\cite{MRHLM21} gives a big improvement than when using simply the weighted polaritonic surface. Work is in progress to extend the self-consistent XF-based MQC methods to the polaritonic case for practical applications. To account for many cavity modes, a classical Ehrenfest treatment for the dynamics of the photonic system has been explored~\cite{HSRKA19}, and
using the XF with the marginal as the photonic displacement coordinate could explain the general underestimation of photon numbers in this approach~\cite{RHLM22} through comparing the exact potential driving the photons with that underlying the Ehrenfest approach.
A further useful factorization could be $\Psi(\dulr,\dulR,\dulq, t) = \chi(\dulR,\dulq,t)\Phi_{\dulR,\dulq}(\dulr,t)$ since this would yield a TDSE for both the nuclear and photonic systems in which the potential ``exactifies" the concept of cavity-BO surfaces introduced in Ref.~\cite{FRAR17}. Also, in general, when several identical, or non-identical particles compose the subsystems, nested commutators along the lines of the formalism of Ref.~\cite{C15} could be considered. 

{\it Perturbative limit} The XF has been employed to address diverse classes of static and dynamic problems in molecules~\cite{Schild_JPCA2016, Scherrer_PRX2017, Scherrer_JCP2015} and solids~\cite{Gross_PRB2019} manifesting essential, though not strong, non-adiabatic effects. For instance, it has been observed that, in determining vibrational circular dichroism of chiral molecules~\cite{Sebastiani_JCP2016, Sebastiani_JCTC2013, Scherrer_JCP2015} or in computing electronic current densities~\cite{Pohl_JPCA2013, Diestler_JCPA2013,  Schild_JPCA2016}, the BO approximation fails to fully account for electronic effects during the dynamics. Even though non-adiabaticity is not strong in these examples because the system does not evolve close to avoided crossings or conical intersections, including excited-state effects as a perturbation to the BO (unperturbed) state is essential -- and sufficient -- to recover the correct behavior. In the context of XF, the electron-nuclear coupling term of Eq.~(\ref{eq:Uen}) has been treated as a perturbation to the BO Hamiltonian~\cite{Scherrer_JCP2015}, where the nuclear velocity is the small parameter. This results in a vector potential that can be seen as a position-dependent mass correction to the bare nuclear masses and that can be used to correct vibrational harmonic frequencies of hydrogen-based molecules~\cite{Scherrer_PRX2017} or phonon frequencies in solids~\cite{Gross_PRB2019}.

{\it Quantum-mechanical clock} The separation of degrees of freedom at the basis of the stationary version of the XF was used in Ref.~\cite{Schild_PRA2018} to introduce the concept of time in quantum mechanics via the classical limit of the quantum-mechanical clock's degrees of freedom $\dulR$ represented by the marginal wavefunction. A clock-dependent continuity equation for the conditional wavefunction was then derived.

\section{Exact potentials driving dissociation and ionization}
\label{sec:diss_ion}
Returning to the theme of dynamics of non-bound states, the exact TDPES of the XF has proved to be a useful interpretative tool. Consider a laser field causing dissociation and ionization in a diatomic molecule. If we are primarily interested in the dissociation aspect, the original factorization of Eqs.~(\ref{eq:exact_el_td}) -- (\ref{eq:exact_A_td}) might be most helpful for analysis since it provides a rigorous definition of the potentials that drive the nuclear motion. If we are interested in the ionization aspect, we may instead want to utilize the reverse factorization, to study the exact potential driving the electronic system. In this section, we discuss examples of these two potentials, previously studied in Refs.~\cite{AMG10,AMG12} and Refs.~\cite{KAM15,KARM17} respectively, for the case of a one-dimensional (1D) model of H$_2^+$, subject to a laser field. 

The model used in Ref.~\cite{AMG10,AMG12} is from Refs.~\cite{KMS96,CFB98,WICCNCBA98,LKGE02}, and involves one electronic coordinate and one internuclear (relative) coordinate with soft-Coulomb-interactions. We note that the laser does not couple directly to the relative nuclear coordinate: the dissociation is driven solely by the electronic motion. In the XF picture, from the nuclear viewpoint this is represented entirely by the TDPES and the time-dependent vector potential. For one-dimensional problems, we can always pick a gauge in which the vector potential vanishes everywhere, and then the TDPES represents the complete potential driving the nuclear system. This is the gauge chosen in this study~\cite{AMG10,AMG12}, and the TDPES is obtained by inversion of Eq.~(\ref{eq:exact_n_td}). The two-dimensional TDSE for the electron-nuclear wavefunction is first solved, starting in the ground state of the molecule, and then $\chi$ is constructed:  its magnitude is  the nuclear density obtained by integrating the full wavefunction over electronic coordinates and its phase is settled by the condition that $A(R,t) = 0$~\cite{AMG12}. 
 
\begin{figure}[H]
 \begin{center}
\includegraphics[width=\textwidth]{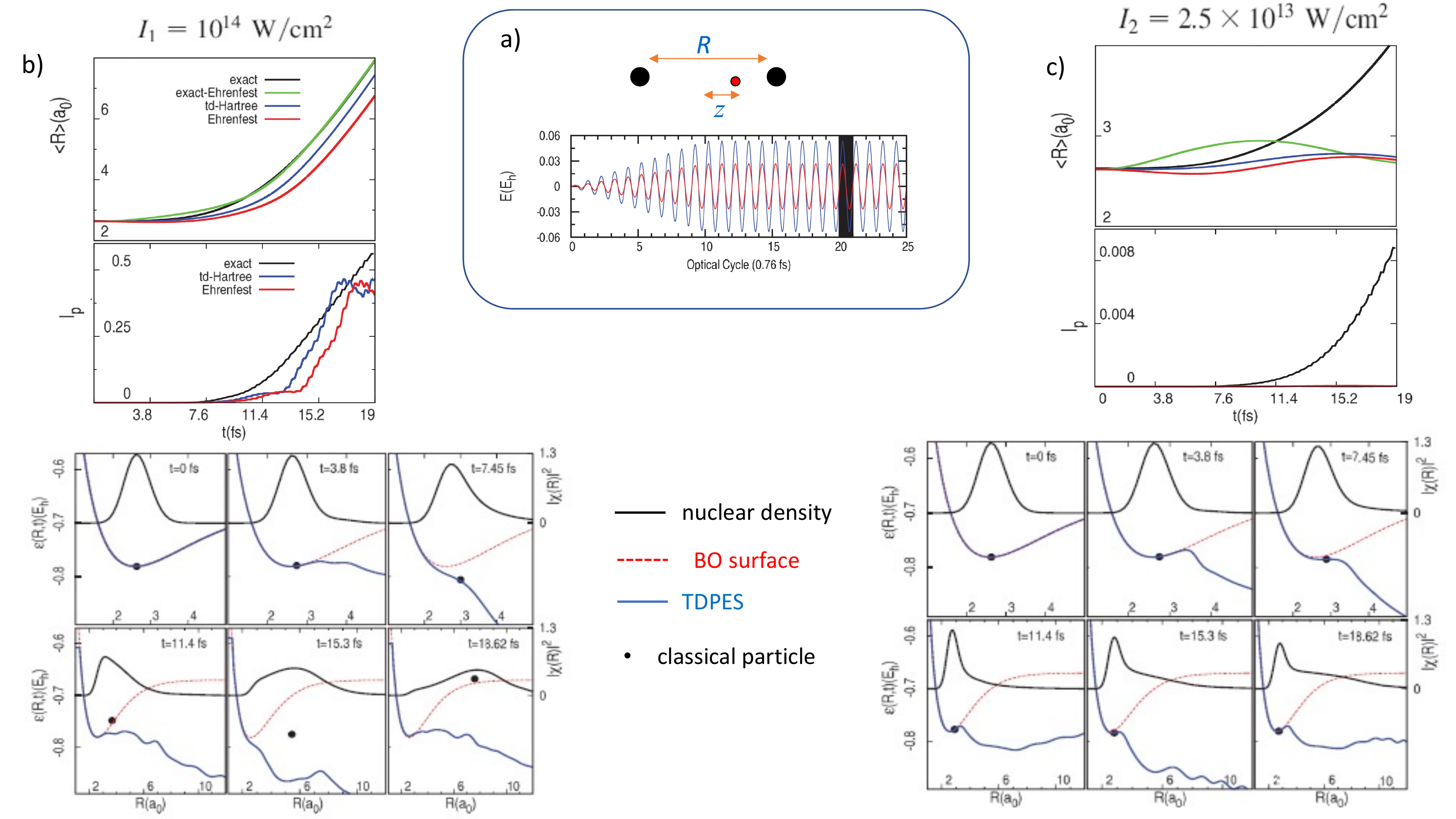}
\caption{1D model of laser-driven H$_2^+$ dissociation: a) Shows a sketch of the model molecule, and the electric field of the $\lambda = 228$nm laser applied to the system, with the shaded part indicating the optical cycle for which snapshots for two different intensities are shown in the other panels.  
b) The top panel shows $\langle R\rangle(t)$ as a measure of dissociation, for the stronger field intensity of $I_1$, and the lower panel shows $I_p = 1 - \int_{\rm box_e} dz\int dR \vert\Psi(z,R,t) \vert^2$ which measures ionization  through the number of electrons outside a box chosen of size $\vert z \vert \le 10$a.u; the exact is shown in black, while predictions of the traditional classical Ehrenfest method is in red, the quantum time-dependent Hartree in blue, and a classical evolution on the exact TDPES (labelled as exact-Ehrenfest) as green. 
The lowest panel shows time-snapshots of the nuclear density and the TDPES, and the black dot shows the position and energy of a classical particle evolving under the TDPES. 
c) This  shows the same quantities as in b) but for the weaker field intensity $I_2$.
Reproduced from Ref.~\cite{AMG12}, https://doi.org/10.1063/1.4745836, with the permission of AIP Publishing. }
\label{fig:diss}
\end{center}
\end{figure}

The laser field  has a wavelength of 228nm, which corresponds to a frequency of about twice the dissociation energy. A sketch is shown in the top middle panel of Fig.~\ref{fig:diss}. We see from the black curves in the upper left panel (b),  that for an intensity of $I_1 = 10^{14}$ W/cm$^2$, the system dissociates, accompanied by significant ionization. This suggests a Coulomb explosion mechanism. On the other hand, for the weaker intensity of $I_2 = 2.5 \times 10^{13}$ W/cm$^2$, over the same time period there is dissociation with $\langle R\rangle$ reaching about half the distance as for the stronger field, but very little ionization (0.008 of an electron by the end of the time shown, as opposed to a little more than 0.5 for the stronger field). The other curves in these plots show the predictions from approximate methods: Ehrenfest,  time-dependent Hartree, and exact-Ehrenfest. In the Ehrenfest approach~\cite{M64,T98}, a single nuclear classical trajectory, initially located at the $\langle R\rangle$ of the initial nuclear probability density and with zero momentum, is propagated by the classical Hamilton's equations with a force given by the average of the soft-Coulomb nuclear-nuclear interaction and electron-density-weighted electron-nuclear interaction. In exact-Ehrenfest, on the other hand, a classical nuclear trajectory is propagated under the gradient of the exact TDPES~\cite{AMG10,AMG12}. These methods both treat the nuclear dynamics classically, while time-dependent Hartree (time-dependent self-consistent field) involves a quantum propagation of a nuclear wavefunction based on an uncorrelated factorization of the molecular wavefunction~\cite{T98} resulting in the  potential driving the nuclei determined self-consistently through the electronic density. 
We observe that while all three of the approximate methods predict dissociation in the stronger field $I_1$, with exact-Ehrenfest being closest to the exact result, none of them is able to predict the dissociation in the weaker field $I_2$ case. The exact TDPES can help us understand why.

The TDPES, shown in the lower set of panels in Fig.~\ref{fig:diss}b and ~\ref{fig:diss}c, shows a strong deviation from the BO surface, with large oscillations in time in the region from near the equilibrium distance on out. The tail of the TDPES alternately falls sharply and returns in correspondence
with the field, which softens the bond, releasing the nuclear density. In this way the TDPES mediates the transfer of energy from the field-accelerated electronic motion to the nuclei. 
The features are similar for both the stronger and weaker field cases, and why a {\it classical} particle evolving on the TDPES does not therefore dissociate for the case of $I_2$ is perhaps not immediately obvious. The answer is revealed when we take a closer look at the structure of the surfaces close to the equilibrium distance and plot the classical energy of the particle evolving in the TDPES (the exact-Ehrenfest calculation) as a function of its position. These are black dots shown in the time snapshots. While for the stronger field $I_1$, the particle quickly becomes loose of the binding potential, being kicked up by the TDPES oscillations, in the case of the weaker field it gets trapped in a narrow potential well not so far from the equilibrium bond length. Without tunneling, it cannot escape, and therefore dissociation does not occur in the classical calculation. This shows that the mechanism for dissociation of the quantum system in the weaker field is tunneling: the TDPES transfers field energy to the nuclei via tunneling, and although the classical exact-Ehrenfest calculation shows a larger amplitude of oscillation of $\langle R\rangle$ than the other approximation methods, it ultimately cannot tunnel through the barrier. Despite treating the nuclei quantum mechanically, so in principle being able to capture effects like tunneling, the time-dependent Hartree method still fails. This is because of its uncorrelated nature:  the electronic density determining the potential for the nuclear motion is  the same for every $R$, and does not deviate far from the exact electronic density near the equilibrium bond-length, due to the weak field strength. The reader is referred to Ref.~\cite{AMG12} where the Hartree surfaces are explicitly shown. This example emphasizes that the mechanism of dissociation via tunneling requires not only an quantum mechanical description of the nuclei but also an adequate description of electron-nuclear correlation. 

We now briefly turn to the electron's viewpoint, and ask what is the potential that is driving its motion? In this case, we choose the reverse factorization where the marginal has the electronic coordinate: $\Psi(\dulr,\dulR,t) = \Phi_\dulr(\dulR,t)\chi(\dulr,t)$, so that the equations are exactly the same as Eq.~(\ref{eq:exact_el_td}) -- (\ref{eq:exact_A_td}) but with $\dulr \leftrightarrow \dulR$. The electronic wavefunction satisfies a TDSE, and, again picking a gauge where $\bA = 0$, the TDPES appearing in the equation for $\Phi(\dulr,t)$ (e-TDPES) together with the external laser field account completely for the electronic system's motion. 

Although formally the same as the direct factorization equations Eq.~(\ref{eq:exact_el_td}) -- (\ref{eq:exact_A_td}), the structure of the TDPES is significantly different, with different terms in Eq.~(\ref{eq:exact_eps_td}) contributing in different ways. In particular,  
the term in the e-TDPES that comes from the electron-nuclear coupling operator $\hat{U}_{en}$, $\frac{1}{2m_e}\langle \Phi_\dulr \vert \nabla_{j}^2 \Phi_\dulr\rangle_\dulR$  (where $\nabla_j$ is the gradient with respect to the $j$th electronic coordinate) is significantly larger in the reverse factorization than in the direct factorization, due to the division by the much smaller electron mass. 
This term tends to give peak structures in the e-TDPES, which can have a significant impact on the electron dynamics. For example, in the case of laser-controlled electron localization~\cite{KSIV11,Sansone2010,HRB07}, mentioned in Sec.~\ref{sec:adventures}, the peak led to an interatomic barrier that enhanced the one in the traditional pictures that comes from Coulomb interaction, resulting in an earlier localization time and smaller localization asymmetry than predicted by traditional methods~\cite{SAMYG14}. The traditional methods use a  ``quasistatic" picture where the electron-nuclear electrostatic potential is added to the laser field to determine the electronic wavefunction's evolution. This work showed that the quasistatic approach misses important effects coming from dynamical electron-nuclear correlation, that can have a notable influence on the predicted observables. 

Dynamical electron-nuclear correlation was also shown to be important in understanding the process of charge-resonance enhanced ionization (CREI). CREI refers to the phenomenon that the ionization rate from a molecule can be several orders of magnitude higher than the rate from the constituent atoms at a critical range of internuclear separations~\cite{ZB95,CB95,CFB98,CZAB95,SIC95,BL12}. This was originally explained by a quasistatic argument, where nuclei are instantaneously fixed point particles. The combined electrostatic electron-nuclear attraction, together with the field, leads to Stark shifted atomic levels, 
and the potential acting on the electrons involves both an outer barrier on the down-field side, and an internuclear barrier that grows as internuclear separation increases. 
Provided
that the field is turned on fast enough such that  population in
the up-field level remains and does not tunnel back to the
down-field atomic level, the molecule can rapidly ionize
over both the inner and outer barriers, at an optimal internuclear separation.  This gives rise to the enhanced
ionization rate. By requiring that the Stark-shifted up-field atomic level exceeds the top of both the inner and outer field modified
Coulombic barriers, one finds the critical
internuclear separation for CREI as 4.07 divided by the ionization potential. The analysis can be generalized
to the case of a laser field whose  period is shorter than
the tunneling time~\cite{ZCB93,ZB95,TB10,TB11}. 

This physical picture is very instructive, but in modeling the experiment, the premise of a quasi-static picture is questionable. A time-resolved study showed that even with static nuclei, the ionization tends to occur in multiple sub-cycle bursts~\cite{TB10,TB11}, not at the peak of the field cycles, which was assumed in the prior  analysis. Furthermore, to observe CREI, the molecule needs to be stretched to the CREI region rapidly enough that appreciable electron density remains un-ionized. In fact, in some experiments, CREI is not observed because there is too much ionization before the critical distance is reached~\cite{BSC06,Bocharova11,Legare03}. Typically, a fraction of the nuclear density remains near equilibrium separations, while a fraction dissociates, and it is the electron density associated with the dissociating fragment that is responsible for CREI. Thus not only the  point particle picture for the nuclei is not accurate, but also the nuclear dynamics means the quasistatic picture is not appropriate, especially as in some dissociating channels the fragment velocities can be comparable to the electronic velocities. Ref.~\cite{KAM15,KARM17} showed that the dynamical electron-nuclear correlation terms in the exact e-TDPES are needed to accurately predict the ionization characteristics when modeling a real CREI experiment. Going beyond the quasistatic treatment
by only accounting for the width and splitting of the
nuclear wave packet is generally not enough to get the
correct dynamics of CREI.

These examples show that the exact TDPES for nuclei or e-TDPES for electrons can provide an illuminating picture on the nature of dynamical electron-nuclear correlation in laser-driven dissociation and ionization, but such studies can only be done when the full molecular TDSE can be somehow solved exactly or highly accurately. 
In the next section, we discuss approximations based on the XF, which move the XF to be a useful approach for real systems.

\section{XF-Based Mixed Quantum-Classical Methods}
\label{sec:MQC}
Being an exact reformulation of the molecular TDSE, the XF equations are no easier to solve than the original molecular TDSE; in fact, due to numerical stability issues, they appear to be harder~\cite{GLM19}! Although the form of the exact coupling terms can give us insight and help us interpret phenomena related to electron-nuclear correlation, to have a practical impact, the XF requires approximations.  MQC approximations are a natural direction for this, justified by the large mass of the nuclei~\cite{EA16}. 

By taking a classical limit for the nuclear motion, and discarding terms in the  equations justified by the exact studies to have a small effect~\cite{AASMMG15}, Refs.~\cite{MAG15,AMAG16} derived the coupled-trajectory MQC (CTMQC) approach.  In CTMQC the electrons are propagated quantum-mechanically whereas the nuclear dynamics consists of classical trajectories coupled through the nuclear quantum momentum
\begin{align}\label{eqn: qmom}
{\bold{Q}_\nu(\dulR,t) = -\frac{\nabla_\nu \vert\chi(\dulR,t)\vert}{\vert\chi(\dulR,t)\vert} = -\frac{\nabla_\nu \vert\chi(\dulR,t)\vert^2}{2\vert\chi(\dulR,t)\vert^2}}
\end{align}
When writing the nuclear wavefunction in terms of its modulus $|\chi(\dulR,t)|$ and phase $S(\dulR,t)$, the term in parenthesis in Eq.~(\ref{eq:Uen}) that depends on $\chi(\dulR,t)$ can be rewritten as
\begin{align}
\frac{-i\nabla_\nu\chi(\dulR,t)}{\chi(\dulR,t)} + \mathbf A_\nu(\dulR,t) = \nabla_\nu S(\dulR,t) +\mathbf A_\nu(\dulR,t) +i\bold{Q}_\nu(\dulR,t)
\end{align}
The first two terms on the right-hand side sum up to the nuclear momentum field, which has a clear classical interpretation. The last term, {\it i.e.} the quantum momentum, appears as an imaginary correction to the classical momentum, with no classical counterpart.

The conditional electronic wavefunction associated with each trajectory $\alpha$ is expanded in the BO basis $\{\Phi^l_\dulR(\dulr)\}$ such that $\Phi_\dulR(\dulr,t) \to \Phi_{\dulR^{\alpha}(t)}(\dulr, t) = \sum_l C_l^\alpha(t) \Phi_{\dulR^\alpha(t)}^l(\dulr)$. Then
the CTMQC equations for the electronic and nuclear evolution are
\begin{equation}
\dot{C}_l^{(\alpha)}(t)=\dot{C}_{l,\textit{Eh}}^{(\alpha)}(t)+\dot{C}_{l,\textit{XF}}^{(\alpha)}(t)\label{eq:ecoef}
\end{equation}
\begin{equation}
\bold{F}_\nu^{(\alpha)}(t)=\bold{F}_{\nu,\textit{Eh}}^{(\alpha)}(t)+\bold{F}_{\nu,\textit{XF}}^{(\alpha)}(t)
\label{eq:force}
\end{equation}
where we note the shorthand $f({\dulR}^{(\alpha)}(t),t)=f^{(\alpha)}(t)$. The first terms in the electronic and nuclear equations are Ehrenfest-like terms
\begin{equation}
\dot{C}_{l,\textit{Eh}}^{(\alpha)}(t)=-\frac{i}{\hbar}E_l^{(\alpha)}C_l^{(\alpha)}(t)-\sum_k\sum_\nu^{N_n}\dot{\bold{R}}_\nu^{(\alpha)}(t)\bold{d}_{\nu,lk}^{(\alpha)}C_k^{(\alpha)}(t)
\end{equation}
\begin{equation}
\bold{F}_{\nu,\textit{Eh}}^{(\alpha)}(t)=\sum_l|C_l^{(\alpha)}(t)|^2(-\nabla_\nu E_l^{(\alpha)})+\sum_{l,k}C_l^{(\alpha)*}(t)C_k^{(\alpha)}(t)(E_l^{(\alpha)}-E_k^{(\alpha)})\bold{d}_{\nu,lk}^{(\alpha)}
\end{equation}
with ${\bf d}_{\nu,lk}^{(\alpha)}$ being the non-adiabatic coupling vector (NAC) along the $\nu$'th nuclear coordinate between BO states $l$ and $k$ evaluated at the coordinate ${\dulR}^{\alpha}(t)$, {\it i.e.} $\langle\Phi_{{\dulR}}^l({\dulr})|\nabla_\nu\Phi_{\dulR}^k({\dulr})\rangle_{{\dulR}^{\alpha}}$ and $E_k^{(\alpha)}$ the $k$-th BO electronic surface.
The second terms in Eqs.~(\ref{eq:ecoef}) and~(\ref{eq:force}) are the corrections coming from XF :
\begin{equation}
\dot{C}_{l,\textit{XF}}^{(\alpha)}(t)=\sum_\nu^{N_n}\frac{\bold{Q}_\nu^{(\alpha)}(t)}{\hbar M_\nu}\left(\bold{f}_{\nu,l}^{(\alpha)}-\sum_k|C_k^{(\alpha)}|^2\bold{f}_{\nu,k}^{(\alpha)}(t)\right)C_l^{(\alpha)}(t)  
\label{eq:ctmqc-e}
\end{equation}
\begin{equation}
\bold{F}_{v,\textit{XF}}^{(\alpha)}(t)=\sum_\mu\frac{2\bold{Q}_\mu^{(\alpha)}(t)}{\hbar M_\mu}\left(\sum_l|C_l^{(\alpha)}(t)|^2\bold{f}_{\mu,l}^{(\alpha)} \right)\left(\bold{f}_{\nu,l}^{(\alpha)}-\sum_k|C_k^{(\alpha)}|^2\bold{f}_{\nu,k}^{(\alpha)}(t)\right)
\end{equation}
The term $\bold{f}_{\nu,l}^{(\alpha)}$ is a momentum that equals the time-integrated adiabatic force accumulated on the $l$th surface  along the trajectory $\alpha$:
\begin{equation}
\bold{f}_{\nu,l}^{(\alpha)}=-\int_{0}^t\nabla_\nu E_l^{(\alpha)} d\tau
\end{equation}
and $\bold{Q}_\nu^{(\alpha)}(t)$ is the quantum momentum evaluated at the position of the trajectory $\dulR^\alpha(t)$
\begin{equation}
\bold{Q}_\nu^{(\alpha)}(t)=-\frac{\nabla_\nu|\chi(\dulR^{(\alpha)},t)|^2}{2|\chi({\dulR}^{(\alpha)},t)|}\label{eq:qmdef}
\end{equation}
In a trajectory-based scheme, at any given time, information of the position of all trajectories is required to reconstruct the nuclear spatial distribution and compute the quantum momentum. Hence, this is a coupled-trajectory scheme, not an independent-trajectory one. 
CTMQC has been successfully employed to simulate several excited-state processes, such as the ring-opening process of oxirane~\cite{MATG17, Tavernelli_EPJB2018} and the photo-isomerization of 2-cis-penta-2,4-dienimiun cation, a retinal photoreceptor~\cite{MOLA20}. Key to its success in these problems is its ability to capture decoherence from first principles, unlike traditional MQC methods as we will shortly discuss.
 
 Another approach is to use the electronic equation derived from XF, i.e. Eq.~(\ref{eq:ecoef}), in a surface-hopping scheme~\cite{HLM18,PyUNIxMD} where the nuclei evolve on a single BO surface at any time, making hops between them according to a stochastic algorithm. This XF-based surface-hopping (SHXF)  has been applied to a range of complex systems, from organic molecules to molecular motors \cite{FMK19,FPMK18,FPMC19,FMC19,VIHMCM21,VMM22}. Part of what has made SHXF able to treat such large systems, is that the quantum momentum term is computed via auxiliary trajectories thus enabling an independent trajectory scheme; these auxiliary trajectories approximately mimic the local coupling of a trajectory to nearby ones. Instead, computing the quantum momentum using coupled trajectories in the same way as in CTMQC, but within a surface-hopping scheme, yields the coupled-trajectory surface-hopping (CTSH) method~\cite{PA21}. More details on the different ways to compute the quantum momentum are given in Sec.~\ref{sec:qmom}, and an investigation of their effect on dynamics in Sec.~\ref{sec:example}. 
  
 In traditional MQC schemes such as Ehrenfest and surface-hopping~\cite{T90,T98,AC19,CB18}, the electronic system suffers from ``overcoherence": after passing through a region of non-adiabatic interactions, the nuclear wavepacket splits with different parts of the wavepacket being correlated with different electronic surfaces. However, the Ehrenfest nuclear distribution follows a mean-field surface and does not split (even if it approximates averaged observables adequately), while, although the surface-hopping nuclear distribution is able to split with different branches evolving on different electronic surfaces, the electronic coefficients remain coherent. In a sense, surface-hopping is more unsettling than Ehrenfest, because of the disconnect between electronic and nuclear systems. 
 Various {\it ad hoc} decoherence corrections have been imposed on the surface-hopping algorithm~\cite{WAP16,CB18,SJLP16}, most often involving an empirical parameter; they are often adequate to reproduce experimental results, but not always, and not reliably. 

The quantum momentum term in Eq.~(\ref{eq:ecoef}) has been shown to give a first-principles description of decoherence in a number of complex systems as well as in detailed studies on model systems~\cite{AMAG16,MATG17,MOLA20,HLM18,GAM18, Agostini_EPJB2018, VIHMCM21}.  But its role goes beyond just decoherence. It was recently shown to be crucial to describe processes where multiple electronic states are simultaneously coupled in some regions of the nuclear configuration space via non-adiabatic coupling~\cite{VMM22}.

\subsection{Computation of the Quantum Momentum}
\label{sec:qmom}
Currently, available codes to perform CTMQC simulations are  G-CTMQC~\cite{GCTMQC,PA21} and PyUNIxMD \cite{PyUNIxMD}. G-CTMQC includes the coupled-trajectory schemes CTMQC and CTSH, with hopping calculated either via fewest-switches hopping, or Landau-Zener, and it is interfaced with the potential library ModelLib~\cite{ModelLib} Furthermore, it has been extended to treat spin-forbidden electronic transitions mediated by spin-orbit coupling, both in the spin-diabatic and spin-adiabatic representations \cite{TMRLA20,TSRLA20}. 
PyUNIxMD has CTMQC and SHXF (with fewest switches hopping probability) capabilities, and is interfaced with a number of electronic structure codes. 
Recently, the simulation of photo-isomerization dynamics of a protonated Schiff base was calculated using both SHXF and CTMQC, achieving results that are in good agreement with each other \cite{KHM22}.

In the CTMQC algorithm there are two main ways to compute the quantum momentum. The first is based on the idea of reconstructing the nuclear density using a sum of gaussians centered at the position of each classical trajectory
\begin{equation}
|\chi^{(\alpha)}(t)|^2= \frac{1}{N_{tr}}\sum_{\beta}^{N_{tr}}{\prod}_{\nu=1}^{N_n}g_{{\boldsymbol\sigma}_\nu^{(\beta)}}(\bold{R}_\nu^{(\alpha)}(t)-\bold{R}_\nu^{(\beta)}(t))
\end{equation}
Note that we have indicated with the symbol $g_{{\boldsymbol\sigma}_\nu^{(\beta)}}(\bold{R}_\nu^{(\alpha)}(t)-\bold{R}_\nu^{(\beta)}(t))$ a normalized three-dimensional gaussian (for each nucleus $\nu$) centered at $\bold{R}_\nu^{(\beta)}(t)$ and with variance $\boldsymbol\sigma_\nu^{(\beta)}$. Then, the quantum momentum is computed analytically  applying Eq.~(\ref{eq:qmdef}).
In the original CTMQC implementation \cite{MAG15,AMAG16,MATG17} the variance of the gaussians was time-dependent and determined from the distribution of classical trajectories along the dynamics. 
However, to stabilize the dynamics and for practical reasons, a frozen gaussian approach with time-independent widths is currently used. This results in the following expression for the quantum momentum:
\begin{equation}
\bold{Q}_\nu^{(\alpha)}(t)=\Gamma_\nu^{(\alpha)}\left(\bold{R}_\nu^{(\alpha)}(t)-\pmb{\mathscr{R}}_\nu^{0\,(\alpha)}(t)\right)\label{eq:qma}
\end{equation}
where the slope of the quantum momentum is computed as $\Gamma_\nu^{(\alpha)}=\sum_\beta W_\nu^{\alpha\beta}$ where
\begin{equation}
W_\nu^{\alpha\beta}=\frac{\prod_{\nu}^{N_n}g_{{\boldsymbol\sigma}_{\nu}^{(\beta)}}(\bold{R}_{\nu}^{(\alpha)}(t)-\bold{R}_\nu^{(\beta)}(t))}{2{\boldsymbol\sigma}_\nu^{(\beta)\,2}\sum_{\beta^\prime}^{N_{tr}}\prod_{\nu^\prime}^{N_n}g_{{\boldsymbol\sigma}_{\nu^\prime}^{(\beta^\prime)}}\left(\bold{R}_{\nu^\prime}^{(\alpha)}(t)-\bold{R}_\nu^{(\beta^\prime)}(t)\right)}\,,
\end{equation}
and the center of the quantum momentum is defined as
\begin{equation}
\pmb{\mathscr{R}}_\nu^{0\,(\alpha)}(t)=\frac{\sum_{\beta^\prime}^{N_{tr}}W_\nu^{\alpha\beta^\prime}\bold{R}_\nu^{(\beta^\prime)}(t)}{\sum_{\beta}W_\nu^{\alpha\beta}}\label{eq:qmca}
\end{equation}
Note that if, for each nucleus $\nu$, all the gaussians carry the same time-independent width ${\boldsymbol\sigma_{\nu}}$, given {\it e.g.} by the widths of the initial nuclear wavepacket, then the slope becomes independent of the trajectory index:
\begin{equation}
\Gamma_\nu^{(\alpha)}\rightarrow\Gamma_\nu=\frac{{1}}{2{\boldsymbol\sigma_{\nu}}^2}
\end{equation}
Because of the approximations introduced to derive CTMQC, the expression for the quantum momentum Eq.~(\ref{eq:qma}) does not satisfy the physical condition that no population transfer among electronic states should be observed in regions where the NAC vectors are zero, once 
averaged over all trajectories. 
The CTMQC equation for the time-variation of the trajectory-averaged population of the $l$th electronic state, $P_l(t) = {N_{tr}^{-1}}\sum_\alpha \vert C_l^{(\alpha)}(t)\vert^2$ is:
\ben
\label{eq:overallpop}
\frac{d P_l(t)}{dt} ={\frac{1}{N_{tr}}} \sum_{\nu\alpha}\frac{2}{M_\nu}\left(\sum_{k}Re(\rho^{(\alpha)}_{lk}){\bf d}_{\nu,lk}^{(\alpha)}{\cdot}\dot{{\bf R}}_\nu^{(\alpha)}+\frac{\bold{Q}_\nu^{(\alpha)}(t)}{\hbar M_\nu}\left(\bold{f}_{\nu,l}^{(\alpha)}-\sum_k\rho_{kk}^{(\alpha)}\bold{f}_{\nu,k}^{(\alpha)}(t)\right)\rho^{(\alpha)}_{ll}(t)\right)
\een
where $\rho^{(\alpha)}_{lk}=C_l^{(\alpha)*}C_k^{(\alpha)}$ are electronic density-matrix elements. To enforce the condition of zero net population transfer when ${\bf d}_{v,lk}^{(\alpha)} = 0$, the second term on the right-hand-side of Eq.~(\ref{eq:overallpop}) is required to be  zero, and this gives an expression for the quantum momentum~\cite{MATG17}. 
Note that this means
the overall population transfer of the whole swarm of trajectories comes solely from the NAC term in Eq.~(\ref{eq:overallpop}) and the quantum momentum affects the electronic population dynamics only indirectly through the evolution of the coefficients appearing in the first term.
The condition is imposed pairwise in the electronic states and for each degree of freedom, which results in a different quantum momentum center for each pair of electronic states 
which is required to be independent of the trajectory index,  $\pmb{\mathscr{R}}_\nu^{0\,(\alpha)}(t)\rightarrow \pmb{\mathscr{R}}_{lk,\nu}^0$: 
\ben
\sum_{\nu\alpha}\frac{2}{M_\nu}\frac{\bold{Q}_{lk,\nu}^{(\alpha)}(t)}{\hbar M_\nu}\left(\bold{f}_{\nu,l}^{(\alpha)}-\bold{f}_{\nu,k}^{(\alpha)}(t)\right)\rho^{(\alpha)}_{ll}(t)\rho^{(\alpha)}_{kk}(t)=0\label{eqn: condition on pop}
\een
Imposing the condition, the numerical expression satisfies
\begin{equation}
\bold{Q}_{lk,\nu}^{(\alpha)}(t)=\Gamma_\nu^{(\alpha)}\left(\bold{R}_\nu^{(\alpha)}(t)-\pmb{\mathscr{R}}_{lk,\nu}^{0}(t)\right)
\end{equation}
where now the center of the quantum momentum is computed as
\begin{equation}
\mathscr{R}_{lk,\nu}^0=\frac{\sum_\alpha\Gamma_\nu^{(\alpha)}R_\nu^{(\alpha)}\left(f_{k,\nu}^{(\alpha)}(t)-f_{l,\nu}^{(\alpha)}(t)\right)|C_l^{(\alpha)}(t)|^2|C_k^{(\alpha)}(t)|^2}{\sum_{\alpha^\prime}{\Gamma}_\nu^{(\alpha^\prime)}\left(f_{k,\nu}^{(\alpha^\prime)}(t)-f_{l,\nu}^{(\alpha^\prime)}(t)\right)|C_l^{(\alpha^\prime)}(t)|^2|C_k^{(\alpha^\prime)}(t)|^2}\label{eq:qmcn}
\end{equation}
A alternative approach is to choose the $y$-intercept of Eq.~(\ref{eq:qma}), $\Gamma_\nu^{(\alpha)}\pmb{\mathscr{R}}_\nu^{0\,(\alpha)}(t)$,  as independent of the trajectory index instead \cite{MATG17}.

In the implementation of the CTMQC algorithm the expression of the quantum momentum center to be used, either the original Eq.~(\ref{eq:qmca}) or the modified Eq.~(\ref{eq:qmcn}), is selected according to a given threshold (referred to as the $\mathcal{M}$-parameter in the code). If the population decoheres and the wavefunction collapses to a single BO state the quantum momentum is set to zero. If the denominator of {Eq.~(\ref{eq:qmcn})} is very small for other reasons different of decoherence, then the analytical expression {(\ref{eq:qmca})} for the center of the quantum momentum is used.

The CTSH algorithm also requires the quantum momentum in the electronic equation, and it is implemented with the same two options as possibilities. 
 
Instead, in the independent-trajectory surface-hopping scheme SHXF,  auxiliary trajectories are used to compute the quantum momentum. In the SHXF algorithm, when the propagated nuclear trajectory enters a region of non-zero non-adiabatic coupling an auxiliary trajectory is created on the BO state in which the population appears. This virtual trajectory propagates with a momentum given by isotropic energy conservation until the BO population in that state becomes negligible and the auxiliary trajectory is killed. At each time step, the quantum momentum is computed placing a gaussian centered at the position of each trajectory and applying the analytical definition Eq.~(\ref{eq:qmdef})
\begin{equation}
\bold{Q}_\nu^{(\alpha)}{(t)}=-\frac{1}{2{\boldsymbol\sigma_{\nu}}^2}\left(\bold{R}_\nu^{(\alpha)}(t)-\sum_l|C_l^{(\alpha)}(t)|^2\bold{R}_{l,\nu}^{(\alpha)}(t)\right)
\label{eq:SHXF}
\end{equation}
 The variance of the gaussian centered on the auxiliary trajectory ${\boldsymbol\sigma_{\nu}}$ in each degree of freedom $\nu$ is an input parameter, and, to eliminate empiricism is chosen to be that of the initial distribution. This has given good results for the real molecules studied. Recently, a new implementation of SHXF~\cite{HM22} was proposed in which a time-dependent width ${\boldsymbol\sigma_\nu}(t)$ is computed by imposing maximum overlap between nuclear wavefunctions on different BO states in phase space.
 
Both SHXF and CTSH follow the same equations, except for the computation of the quantum momentum, and, in particular, Eq.~(\ref{eq:SHXF})  of SHXF approximately mimics the computation with coupled trajectories of CTSH using the original definition Eq.~(\ref{eq:qmca}) in Eq.~(\ref{eq:qma}). Although similar in spirit, we see that there is a significant difference in centers of the quantum momentum (second term on the right of Eq.~(\ref{eq:SHXF}) {\it c.f.} Eq.~(\ref{eq:qma})). For two-state problems, while all trajectories of the ensemble contribute in CTSH,  only two trajectories enter in SHXF: the running state, and one auxiliary trajectory. In fact, in the two-state case, Eq.~(\ref{eq:SHXF}) reduces to
\ben
\bold{Q}_\nu^{(\alpha)}=-\frac{1}{2{\boldsymbol\sigma_{\nu}}^2}\left(\bold{R}_{\nu,a}^{(\alpha)}(t) - \bold{R}_{\nu,in}^{(\alpha)}(t)\right)(1 - \vert {C_a^{(\alpha)}}(t)\vert^2)
\label{eq:shxf_2state}
\een
where $\bold{R}_{\nu,a}^{(\alpha)}(t)$ denotes the position of the active trajectory (running state), with electronic coefficient ${C_a^{(\alpha)}}(t)$, and $\bold{R}_{\nu,in}^{(\alpha)}(t)$ the auxiliary trajectory. In contrast, the center of the quantum momentum in CTSH is in practice vastly different in Eq.~(\ref{eq:qmca}), including contributions from all trajectories on all surfaces, even though the originating equation is the same.

 The main advantages of SHXF are computational efficiency and numerical stability. However, both SHXF and CTSH suffer from deficiencies inherited from the {\it ad hoc} nature of the surface-hopping procedure itself: the ambiguity of velocity-adjustments after a hop~\cite{VIHMCM21,B21,CGB17,TSF21}, and the incorrect forces in the vicinity of a NAC region where the exact force tends to have a diabatic or averaged character rather than that from one surface. 
 Further, although the disconnect between the single-surface evolution of the nuclei and the coherent electronic evolution is partially ameliorated with the XF term, reducing the difference between the electronic population at a given time and the fraction of trajectories evolving on a given surface,
 this so-called ``internal consistency" is not guaranteed. 
Finally, 
from the XF side, we note that the modified computation of the quantum momentum Eq.~(\ref{eq:qmcn}) is not possible within SHXF because the trajectories are not coupled to each other.
\section{Computing the quantum momentum}
\label{sec:example}

We now investigate the effect on the dynamics of a one-dimensional system of the three ways of calculating the quantum momentum term: with coupled-trajectories (CTSH and CTMQC) using the original definition Eq.~(\ref{eq:qmca}), and using the  modified implementation Eq.~(\ref{eq:qmcn}), and with the auxiliary trajectories method (as done in SHXF).
 Further we compare the effect of the XF terms in the nuclear equation of CTMQC versus the surface-hopping schemes SHXF and CTSH.

\begin{figure}[H]
 \begin{center}
\includegraphics[width=\textwidth]{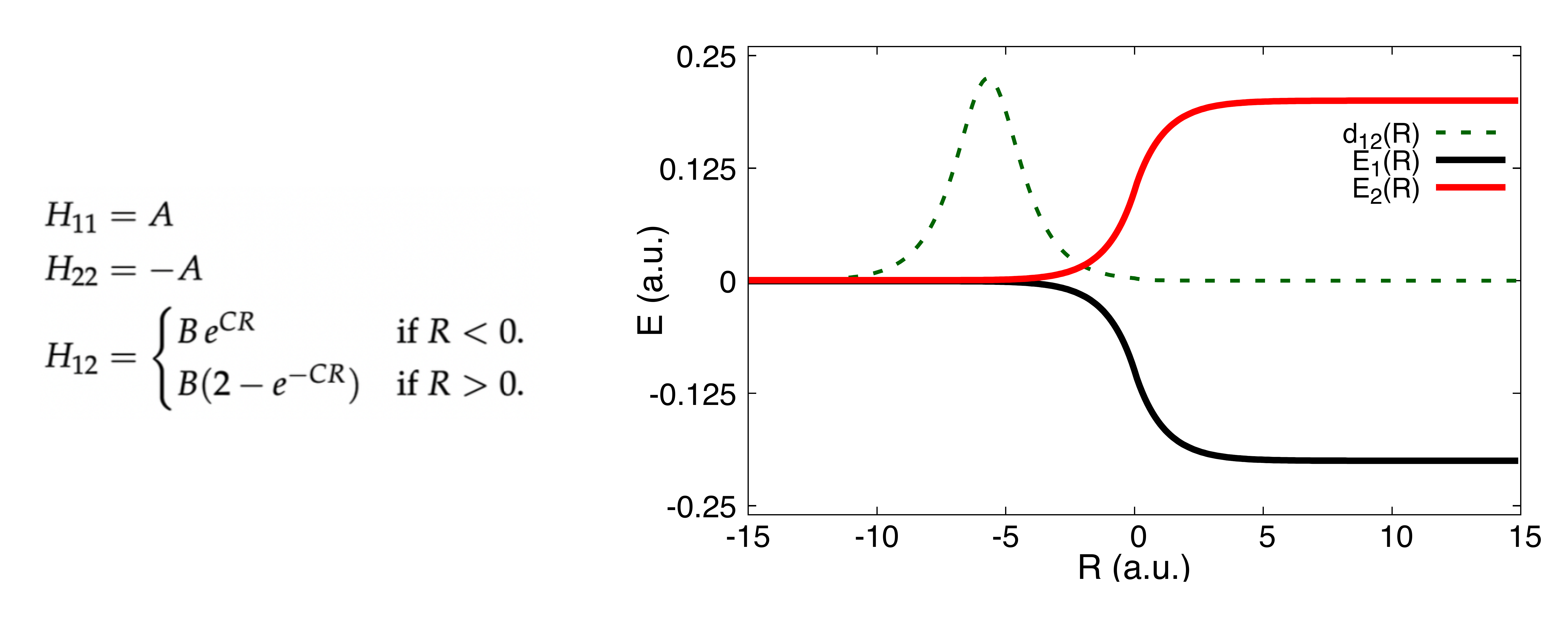}
\caption{Analytical form of the electronic Hamiltonian on the left, with the BO potential energy surfaces (red and black) and NAC (dashed green) shown on the right for ECR model. }
\label{fig:tully}
\end{center}
\end{figure}

We choose our model system as the Tully model with an extended coupling region with reflection (ECR)~\cite{T90}, where an incoming nuclear wavepacket enters a region of extended non-adiabatic coupling between two BO surfaces that are asymptotically parallel. 
Along with other Tully models, the performance of CTMQC was studied in detail in Ref.~\cite{AMAG16}. The analytical form of the electronic Hamiltonian {matrix elements}, the BO surfaces, and the NAC are shown in Fig.~\ref{fig:tully}. 

The parameters were chosen to be $A=6 \times 10^{-4}$~{a.u.},  $B=0.1$~{a.u.} and $C=0.90$~{a.u.}
We will study two cases, a Gaussian nuclear wavepacket incoming on the lower surface from the left with a higher ({$k_0=30$}~a.u.) and a lower ({$k_0=10$}~a.u.) initial momentum. In the  NAC region, some population is transferred to the upper surface and in both cases, the off-diagonal elements of the density matrix in the BO basis (coherences) rise and then eventually vanish after passing the non-adiabatic region: 
For the higher initial momentum, the nuclear wavepacket associated with the lower surface moves faster, separating in nuclear space from the part that has transferred to the upper surface, while for the low initial momentum case 
 one branch of the nuclear wavepacket is reflected and the other is transmitted. The decay of the coherence cannot be accounted for by uncorrected surface hopping since the electronic equations are propagated fully coherently along each independent trajectory~\cite{T90}, and the method becomes ``internally inconsistent" in that BO population determined from the modulus-square of the associated electronic coefficient differs from that determined by the fraction of nuclear trajectories evolving on that surface~\cite{T90,T98,AC19,CB18, WAP16,SJLP16}.

We ran 1000 Wigner-distributed trajectories starting on the lower BO surface for CTSH and SHXF, while CTMQC calculations converge already with 200 trajectories. The initial position was $R_0=-15$~a.u. and the variance $\sigma_0$ of the gaussian wavepacket at time zero was chosen to be 20 times the inverse of the initial momentum. The time-step used in the calculations is 0.1 a.u.

\subsection{ECR model $k_0 = 30${~a.u.}}
\label{sec:highk}

\begin{figure}[h!]
 \begin{center}
\includegraphics[width=\textwidth]{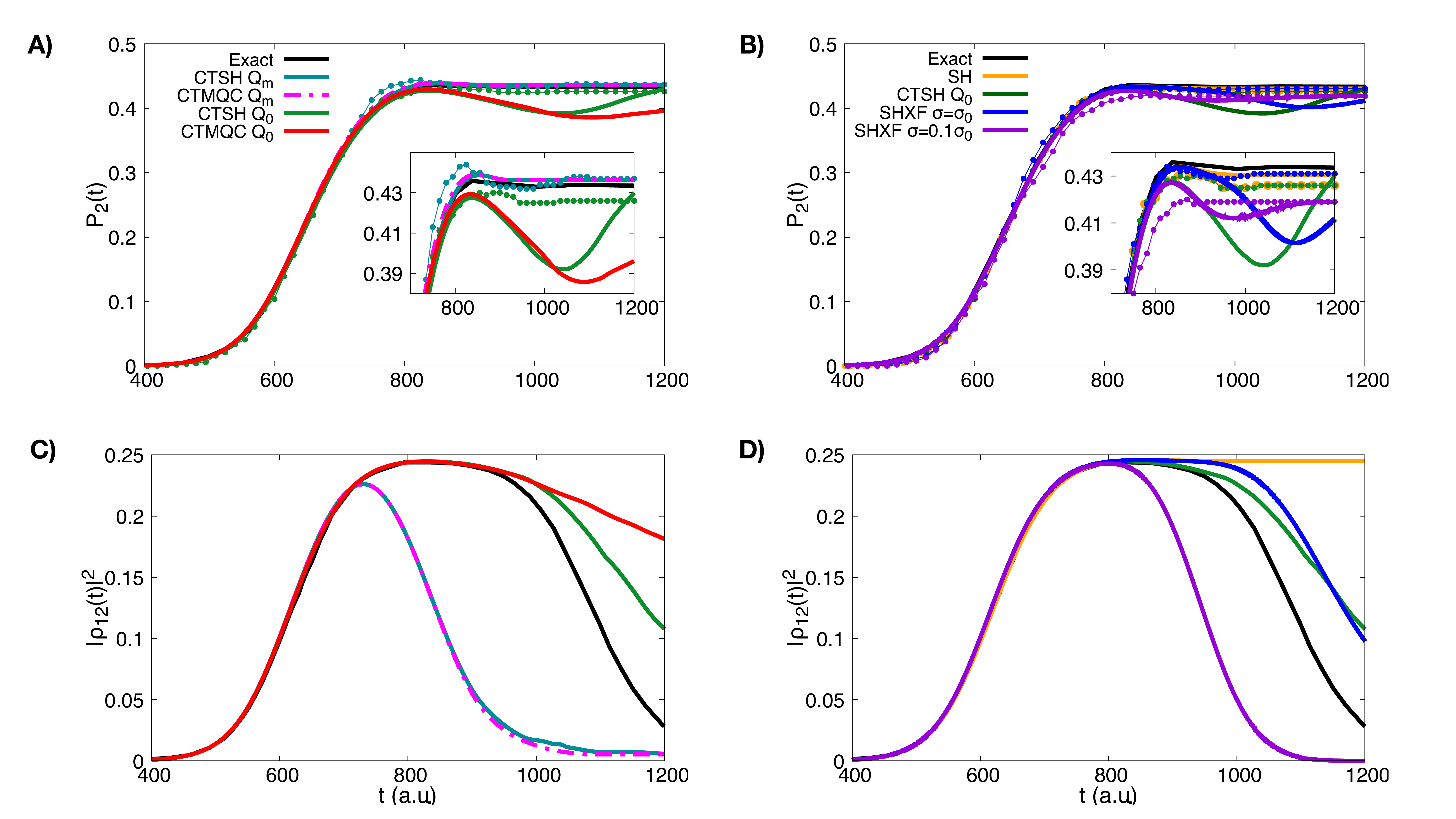}
\caption{
Upper panels: electronic populations {of the excited state} (solid) and fraction of trajectories {on the excited state} (dotted) are plotted for ECR model {with} $k_0=30$ a.u., as functions of time starting at 400~a.u. which is when the trajectories and the quantum wavepacket approach the NAC region. The label ${\rm Q_m}$ indicates the use of Eq.~(\ref{eq:qmcn}), {\it i.e.} the modified definition of the quantum momentum, while ${\rm Q_o}$ indicates the use of Eq.~(\ref{eq:qmca}), {\it i.e.} the original definition. Lower panels: electronic coherences, $\vert\rho_{12}\vert^2 = \sum_\alpha^{N_{traj}} \vert C_1^{\alpha *}(t)C_2^\alpha(t)\vert^2/N_{traj}$,  calculated by averaging over the trajectories and using their expressions in terms of the electronic coefficients. The color code is the same as in the upper panels. The coupled-trajectory results are compared with exact results in the left panels, whereas auxiliary-trajectory results are shown in the right panels. For reference, surface-hopping (SH) results with no decoherence corrections are shown in the right panels. } 
\label{fig:ecr_pop_coh}
\end{center}
\end{figure}   
As we can see from panel A) in Fig.~\ref{fig:ecr_pop_coh}, both CTMQC and the CTSH electronic populations with the modified definition of the quantum momentum (Eq.~(\ref{eq:qmcn})) closely reproduce the exact result. For CTSH, measuring the population instead by the fraction of trajectories evolving on the surface yields small oscillations around the exact results. 
On the other hand, panel A) in Fig.~\ref{fig:ecr_pop_coh} shows that both CTMQC and CTSH with the original definition of the quantum momentum predict a small population transfer back to the ground state after the first interaction region when considering the electronic populations. Interestingly, the fraction of trajectories measure, however, does not show this, and instead has a similar trend as the exact but with an underestimate. Comparing with panel B) of Fig.~\ref{fig:ecr_pop_coh}, we see that the trend in the electronic populations is the same in SHXF. That is, the electronic populations yield a back-transfer when the original definition of the quantum momentum is used, whether it is computed via auxiliary trajectories as in SHXF or with coupled-trajectories as in CTSH. The two ways yield similar results for the populations, although it should be noted that the SHXF results show a dependence on the width-parameter of Eq.~(\ref{eq:SHXF}). Taking $\sigma = \sigma_0$  arguably gives an overall closer result to the parameter-free CTSH population, although SHXF with $\sigma =  \sigma_0/10$ matches closely the CTSH populations at earlier times.
The fraction of trajectories in SHXF does not show the back-transfer, just as that in CTSH did not; this measure in SHXF with $\sigma =  \sigma_0/10$ gives a larger underestimate of the population transfer. But the fraction of trajectories measure for either $\sigma$-parameter in SHXF as well as either definition of the quantum momentum in  CTSH are quite close to the exact, and close to that of CTMQC with the modified definition. At the same time we observe that uncorrected SH electronic populations are close to that of the exact while the fraction of trajectories closely resembles that of CTSH. However the uncorrected SH coherences do not decay in time and would lead to errors in later evolution if the system encounters another interaction region.

\begin{figure}[h!]
 \begin{center}
\includegraphics[width=\textwidth]{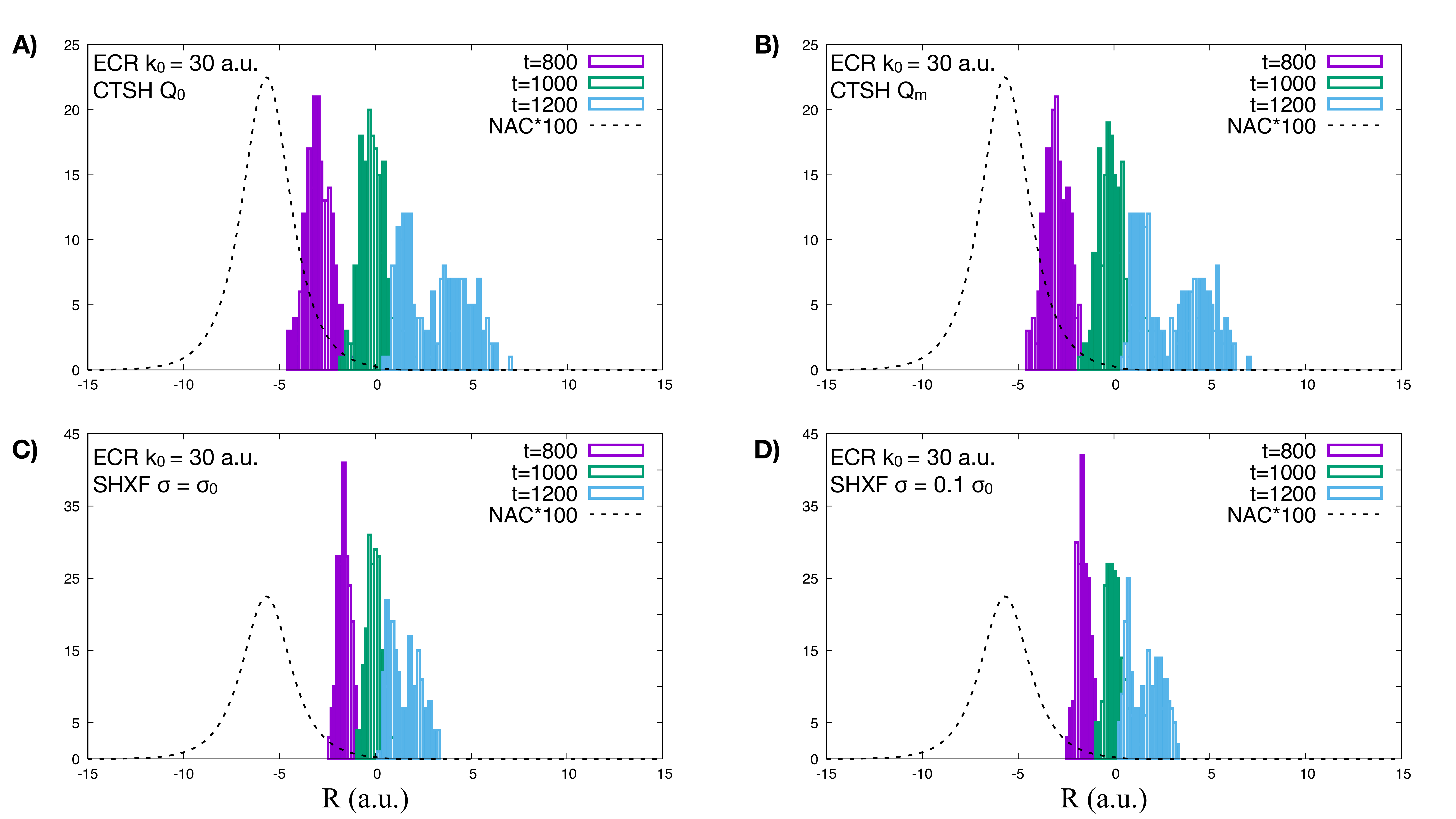}
\caption{Spatial distribution of trajectories for the ECR model with $k_0=30$~a.u.: CTSH with original definition of the {quantum momentum using Eq.~(\ref{eq:qmca})} (panel A), CTSH with the modified definition of the {quantum momentum using Eq.~(\ref{eq:qmcn})} (panel B), SHXF with $\sigma=\sigma_0$ (panel C) and SHXF $\sigma=\sigma_0$/10 (panel D). }
\label{fig:histograms}
\end{center}
\end{figure}

The incorrect back-transfer of population seen in the electronic populations when using the original definition of quantum momentum can be related directly to the violation of the condition of zero net population transfer in regions of zero NAC. To demonstrate this, we plot  the spatial distribution of trajectories at different times in Fig.~\ref{fig:histograms} for CTSH and SHXF along with the NAC. The top panels show that the nuclear distributions for CTSH with the two definitions are similar at these times, and that between 1000 and 1200 a.u., only a small fraction of trajectories are in a region of non-negligible non-adiabatic coupling; so the population transfer seen with the original definition Eq.~(\ref{eq:qmca}) is a violation of the condition. The lower panels show a similar violation occurs for SHXF $\sigma=\sigma_0$ and $\sigma_0/10$, consistent with the unphysical population transfer seen there. However, using the fraction of trajectories appears to remedy this problem: the fewest-switches hopping probability is proportional to the NAC so if this is too small, then the system is very unlikely to hop. 

Thus, in a sense, the fewest-switches hopping probability somewhat takes care of the zero net population transfer in regions of zero NAC, since the probability is zero when the NAC is. We trade one violation for another: 
the violation of the internal inconsistency enables us to obtain physically reasonable populations through the fraction of trajectories measure of populations instead of through the electronic coefficients, which avoids us violating the condition of zero net population transfer in regions of zero NAC. 

Looking at the coherences plotted in panel C) of Fig.~\ref{fig:ecr_pop_coh}, both CTSH and CTMQC with the modified definition of the quantum momentum are in agreement with each other and show faster decoherence rates than the exact calculations while using the original definition of the quantum momentum predicts slower decoherence times. On the other hand, as we see in panel D) the coherences for SHXF $\sigma=\sigma_0$ are similar to those of CTSH with the original quantum momentum definition, and the decoherence rate increases with decreasing $\sigma$ as expected from Eq.~(\ref{eq:SHXF}).

\subsection{ECR model $k_0 = 10$~{a.u.}}
\label{sec:lowk}

\begin{figure}[h!]
 \begin{center}
\includegraphics[width=\textwidth]{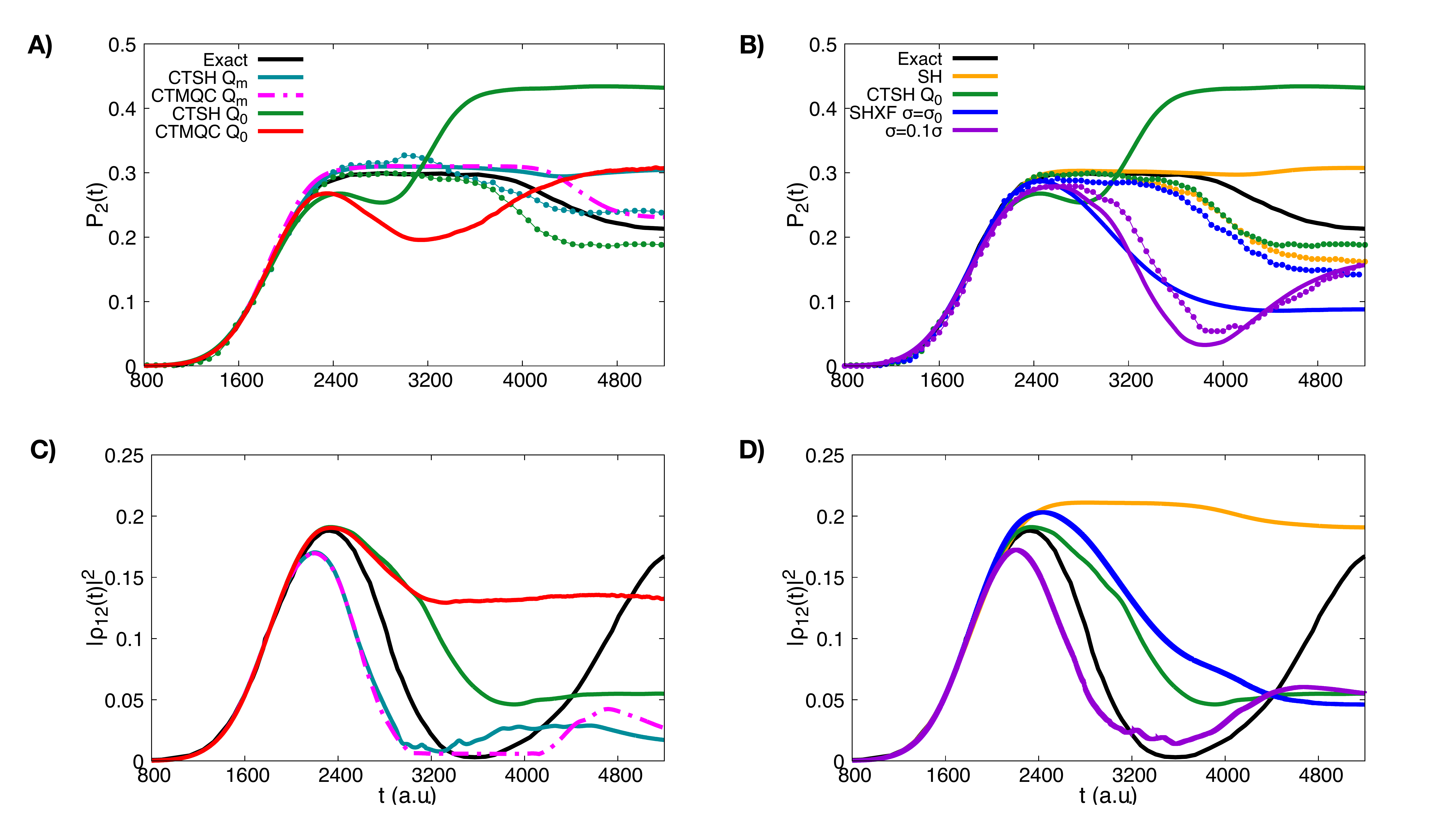}
\caption{Upper panels: electronic populations {of the excited state} (solid) and fraction of trajectories {on the excited state} (dotted) are plotted for ECR model {with} $k_0=10$ a.u. Lower panels: electronic coherences}
\label{fig:k10pops_cohs}
\end{center}
\end{figure}   
 
As we can see in Fig. \ref{fig:k10pops_cohs}, in contrast to the higher momentum case, there is substantial  disagreement between the methods for the low incoming momentum case. 

First let us consider the coupled-trajectory methods with the modified quantum momentum expression, Eq.~(\ref{eq:qmcn}). 
  CTMQC with Eq.~(\ref{eq:qmcn}) closely reproduces the exact results, slightly overestimating the population transfer to the excited state and underestimating the back-transfer from the part of the nuclear wavepacket that reflects. Using Eq.~(\ref{eq:qmcn}) the electronic populations of CTSH however do not reproduce the transfer back to the ground state that would be associated with reflecting trajectories.  On the other hand, the CTSH fraction of trajectories reproduces the populations better than the electronic populations, even though, in a sense the latter are more fundamentally derived~\cite{PA21}; we had seen this also for the higher momentum case in Sec.~\ref{sec:highk} when the original definition of the quantum momentum was used, but now here we see it also for the modified definition. 
Unlike the higher momentum case, this cannot be attributed to violation of the zero net population transfer condition since the modified definition respects this condition.

\begin{figure}
 \begin{center}
\includegraphics[width=0.68\textwidth]{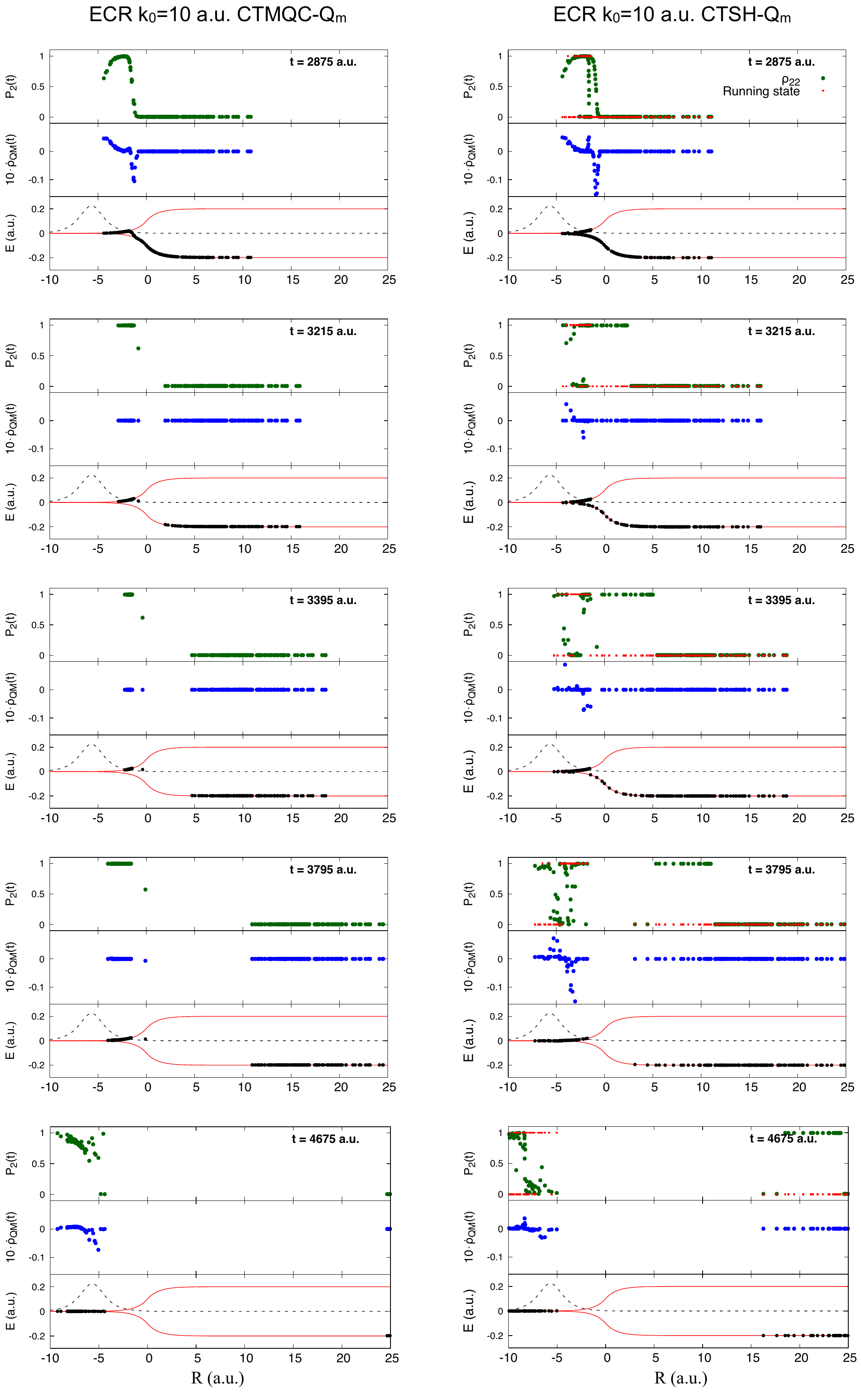}
\caption{Top panels in each plot: Time snapshots of the electronic populations of the excited state for CTMQC (left) and CTSH (right), with the modified definition of the quantum momentum, at the positions of the trajectories (green dots) and running state (red dots). Middle panels in each plot: XF contribution to the time-variation of the electronic populations ({\it i.e.} the second-term on the right-hand-side of Eq.~(\ref{eq:overallpop}) but without the trajectory-sum). Lower panels in each plot: BO surfaces (thin red lines), NAC (thin dashed lines), and spatial distribution of the nuclear trajectories on the BO surfaces.}
\label{fig:coeff_evolk10}
\end{center}
\end{figure}

With the modified definition of the quantum momentum, although the CTSH electronic populations are similar to CTMQC until about 4000 a.u., the coefficients are different from about 3000 a.u. onwards, as is evident from the coherences shown in the lower panel. CTMQC coherences decay after the first passage on a similar but shorter time-scale as the exact before rising later when the reflected trajectories pass the NAC. But CTSH coherences rise after a much shorter time: the CTSH trajectories spend more time in the NAC region, reflecting earlier, and the coefficients of the individual trajectories evolve, even if the population averaged over all trajectories remains constant.  The quantum momentum term (which depends on $\vert C_1^{(\alpha)}(t)\vert^2\vert C_2^{(\alpha)}(t)\vert^2$) in the equation of motion for the coefficients remains activated in CTSH from 3200 a.u. onwards, unlike in CTMQC where it is essentially zero from 3200 a.u. to about 4000 a.u. These observations are illustrated in Fig.~\ref{fig:coeff_evolk10}.  The figure  also shows the electronic energies associated with each trajectory, which, for the CTMQC  maps out the effective approximate TDPES. While the CTSH trajectories dwell in the NAC region, there is a significant hopping probability, yielding changes in which surface a given trajectory evolves on. Even at large times, the internal inconsistency is severe in both the reflected and transmitted region. In particular, in the first non-adiabatic event, some trajectories do not hop but the associated electronic coefficient changes to the upper surface leading to a transmitted part of the nuclear distibution with poor internal consistency as we observe on the trailing edge of the transmitted trajectories (Fig.~\ref{fig:coeff_evolk10}). This is responsible for the rise in the CTSH populations between 3000 and 4000 a.u. Arguably the  trend of the fraction of trajectories measure of populations in Fig.~\ref{fig:k10pops_cohs} is closer to the exact than that from the electronic populations, just as the running state population is closer to that of CTMQC in Fig.~\ref{fig:coeff_evolk10}. 
The reason why the CTSH trajectories stay longer in the NAC region and reflect earlier than the CTMQC trajectories is related to the effective potential energies they experience: the ones which have hopped to the upper surface evolve on the more sharply rising upper BO surface, with a velocity penalty to conserve energy from making the hop, while the CTMQC ones evolve on an approximate TDPES which tends to be bounded by the upper BO surface but can be below it.  
The CTMQC trajectories (black dots moving on the thin red lines representing the BO energy curves in the left panels) leave the NAC region earlier than the CTSH ones (black dots in the right panels), and move further to the right, before the reflection occurs. While at the beginning of the non-adiabatic event (shown at time $t=2875$~a.u. in Fig.~\ref{fig:coeff_evolk10}) the distributions of CTMQC and CTSH trajectories are similar, the issues just described can be identified at later times, for instance at $t=3215$~a.u. and $t=3395$~a.u. in Fig.~\ref{fig:coeff_evolk10}. The distribution generated by the CTSH evolution at $t=3795$~a.u. and $t=4675$~a.u.  has a splitting that agrees better with CTMQC than at previous times. Movies describing the full dynamics as documented in Fig.~\ref{fig:coeff_evolk10} are provided as Supplementary Material.

In Fig.~\ref{fig:coeff_evolk10}, comparing the CTSH electronic populations (green dots)  in the top right-hand plots at each time with the running state (red dots, or, compare with the bottom plots) attests to the breakdown of the internal consistency of the algorithm, discussed above. Specifically, we note that in some regions of space and already at early times, the running state for each trajectory (the red dots are at zero for the ground electronic state and one for the excited state), differs from the value of the excited-state population for those same trajectories. When a trajectory is on the ground-state potential energy curve, the corresponding population associated to the excited state ($\rho_{22}^\alpha(t)$) should be zero if internal consistency holds. However, this is not always the case, as evident from Fig.~\ref{fig:coeff_evolk10}. For CTMQC, the issue of internal consistency does not exist. We show for completeness, in the same figure but in the middle plots of each panel, the contribution of the quantum-momentum XF term in the electronic evolution equation (in particular, we show as blue dots the second term in the right-hand side of Eq.~(\ref{eq:overallpop}) but without the trajectory-sum). We note that at times $t=3215, 3395, 3795$~a.u. the effect of such term is non-zero, in contrast to what is predicted by CTMQC.

\begin{figure*}
 \begin{center}
\includegraphics[width=0.7\textwidth]{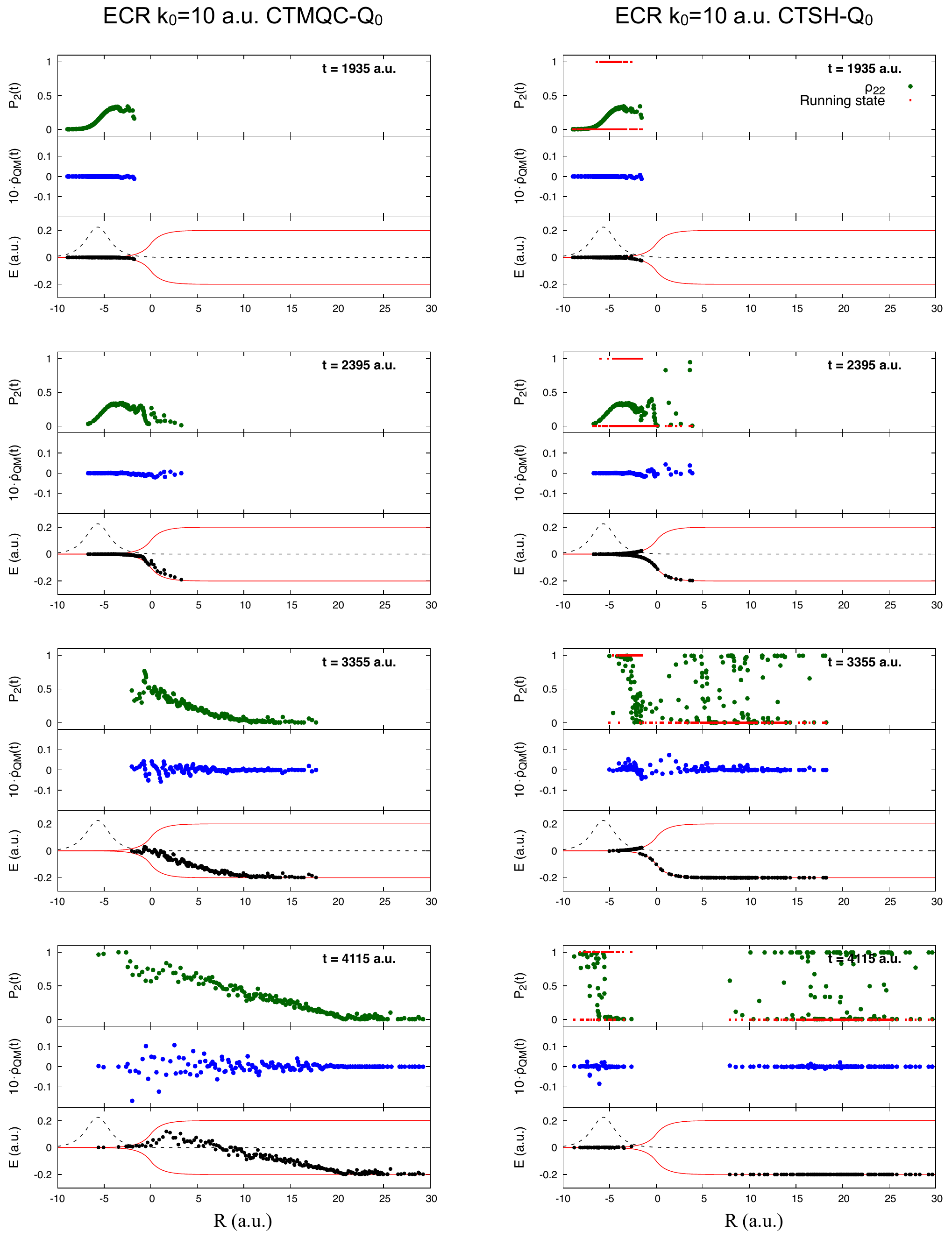}
\caption{Same as in Fig.~\ref{fig:coeff_evolk10} but using the original definition of the quantum momemtum.}
\label{fig:coeff_evol_o}
\end{center}
\end{figure*}

We turn now to the CTMQC and CTSH results using the original definition of the quantum momentum Eq.~(\ref{eq:qmca}). With this definition we see from the lower left panel of Fig.~\ref{fig:k10pops_cohs} that neither of these decohere during this time period. The CTMQC populations continue to evolve throughout, beginning to differ from the exact result already around about 2400 a.u. Soon after this time, many of the trajectories have moved passed the NAC region, as evident in Fig.~\ref{fig:coeff_evol_o} (black dots in the bottom panels of each plot on the left shown at different times along the dynamics -- those times are different than Fig.~\ref{fig:coeff_evolk10} since the overall evolution is different if compared to the previous case where the modified quantum momentum was used), and there is violation of the zero net population transfer condition in this part of the nuclear wavepacket.  In fact the contribution of the quantum momentum term to the equation of motion for the coefficient (blue dots on the left panels in Fig.~\ref{fig:coeff_evol_o}) can be larger in the transmitted part than it is in the NAC region, which is a big distinction with the modified quantum momentum result where it was zero in the transmitted region (compare Fig.~\ref{fig:coeff_evolk10} and~\ref{fig:coeff_evol_o}). 
Driven by the quantum momentum (which is predominantly negative in this leading edge~\cite{GAM18}) populations in this region start transferring back to the ground state after previously having transferred to the excited state, earlier than in when the modified quantum momentum definition is used.  The effective TDPES shown by the electronic energies associated with each CTMQC trajectory shows a significant difference from that when using the modified definition of the quantum momentum in Fig.~\ref{fig:coeff_evolk10}, and does not reduce to the BO surfaces away from the interaction region as they should. The distribution of nuclear trajectories do not show a clear splitting. Finally, note that the population of the electronic excited state at the positions of the trajectories (green dots on the left in Fig.~\ref{fig:coeff_evol_o}) clearly attests to the lack of spatial splitting and efficient decoherence observed for the dynamics of the trajectories.

The electronic populations of CTSH with the original quantum momentum definition Eq.~(\ref{eq:qmca})
show a strikingly different behavior than that of CTMQC, with many continuing to rise after the first transfer event, before leveling of (see the right panels of Fig.~\ref{fig:coeff_evol_o} for CTSH results to be compared to the CTMQC results just described). As in CTMQC with this definition, there is violation of the zero net population transfer condition in the transmitted wavepacket, but the populations behave very differently: each of the transmitted and reflected wavepackets have populations spanning both upper and lower surfaces in CTSH, and the unphysical growth of the trajectory-averaged populations in Fig.~\ref{fig:k10pops_cohs} comes primarily with trajectories associated with the transmitted part of the wavepacket. In contrast, in CTMQC, there is a linear shape to the population going from more on the upper surface on the reflected side to more on the lower surface on the transmitted packet. 
Interestingly, the CTSH fraction of trajectories is very similar to that with the modified definition of the quantum momentum, and has a trend closer to the exact. Hopping can only occur in the region of the NAC, so, as observed in the higher-momentum case, in a sense ameliorates the violation of the condition of zero net transfer in regions of zero NAC. Movies describing the full dynamics as documented in Fig.~\ref{fig:coeff_evol_o} are provided as Supplementary Material.

\begin{figure}[h!]
 \begin{center}
\includegraphics[width=\textwidth]{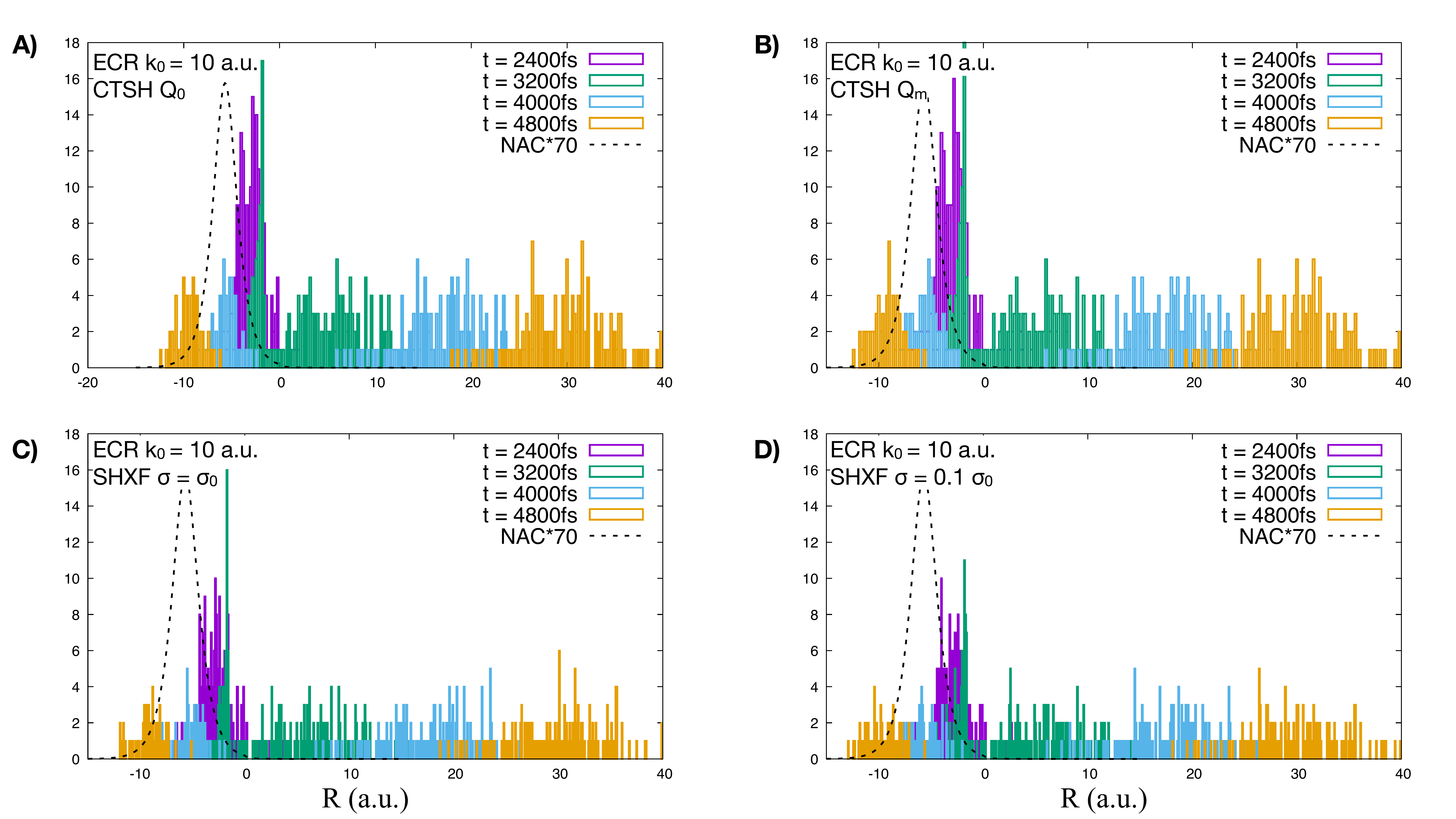}
\caption{Spatial distribution of trajectories for the ECR $k=10$~a.u. model. CTSH with original definition of the QM (upper left panel), CTSH with the modified definition of the QM (upper right panel), SHXF with $\sigma=\sigma_0$ (lower left panel) and SHXF $\sigma=\sigma_0$/10 (lower right panel).}
 \label{fig:spatial_distk10}
\end{center}
\end{figure}   

Turning now to the independent trajectory surface-hopping calculations shown in the upper right panel of Fig.~\ref{fig:k10pops_cohs}. We see that uncorrected surface hopping  in fact looks very similar to CTSH with the modified definition of the quantum momentum, in both the populations as well as the fraction of trajectories but gives poor coherences (lower right panel). 
Including the XF correction with $\sigma = \sigma_0$, leads to a large back-transfer of electronic populations, but again with a similar fraction of trajectories as the uncorrected surface hopping and the CTSH cases. Taking instead $\sigma = \sigma_0/10$ further causes the system to decohere faster initially and improves the internal consistency but at the expense of worse agreement with the exact. A comparison of the histogram of the nuclear trajectories at different time-snapshots is shown in Fig.~\ref{fig:spatial_distk10} for CTSH with the modified and original definitions of quantum momentum and SHXF with the two different values of $\sigma$, showing they yield similar nuclear distributions as a function of time, consistent with the fraction of trajectories being similar. 
At the later times when the transmitted part of the wavepacket has left the NAC region, the associated populations of CTSH Q$_0$ and SHXF however continue to evolve (compare with Fig.~\ref{fig:coeff_evol_o} and the movies), consistent with the violation of the condition of zero net population transfer. 
 
Interestingly, the trend in the populations of SHXF is more similar to that in CTMQC with Eq.~(\ref{eq:qmca}) and opposite to that of CTSH (with Eq.~\ref{eq:qmca}), after the first interaction region. Despite being similar in the underlying equations, this may not be surprising given the very different implementation of Eq.~(\ref{eq:qmca}) in the two methods (see discussion at the end of Sec.~\ref{sec:qmom}). In the case of the higher-momentum $k = 30$ a.u., the population trends in CTSH and SHXF were in fact similar: likely because there is essentially one localized wavepacket on the upper surface and one on the lower, so the coupled-trajectory computation of the quantum momentum with Eq.~(\ref{eq:qmca})  is similar to the auxiliary-trajectory way which takes simply each independent active trajectory plus one auxiliary trajectory on the other surface. 
For the lower momentum case $k = 10$ a.u., two wavepackets develop on the lower surface going in opposite directions, as well as a wavepacket on the upper surface: while CTSH computes the quantum momentum from contributions from all of these, again the SHXF just takes the independent active surface trajectory plus one auxiliary trajectory on the other surface, trajectory-by-trajectory. 

\section{Discussion}
\label{sec:concs}
The XF approach offers new insights into the interaction of coupled quantum subsystems, through its definition of exact coupling terms between the wavefunctions. For the degrees of freedom chosen as the marginal in the factorization, one obtains an effective TDSE in which the potentials contain the full effects of the correlated motion of the other subsystems. This has proven instructive in revealing molecular dissociation mechanisms under different laser parameters, as well as highlighting the effects of electron-nuclear correlation in ionization processes going beyond the usual quasi-static picture of the nuclear potentials. The generality of the XF idea has led to adventures in a wide range of applications outside coupled electron-nuclear dynamics, including electron-electron factorization to define a new single-active electron picture, quantum embedding for strongly correlated electronic systems, and inclusion of photons for molecular polaritonics. 

The most practical impact of the XF approach so far has been in the development of rigorous MQC methods, and CTMQC and SHXF have both successfully predicted dynamics in a number of complex situations~\cite{MATG17, Tavernelli_EPJB2018, MOLA20,FMK19,FPMK18,FPMC19,FMC19,VIHMCM21}, giving improved results from first-principles  compared to traditional MQC methods~\cite{VMM22}. These methods involve computing a term that depends on the nuclear quantum momentum, for which three different implementations exist. In particular, with coupled trajectories, there is the original definition coming directly from $\nabla_\nu \vert\chi(\dulR,t)\vert/\vert\chi(\dulR,t)\vert$, as well as a modified definition that ensures the physical condition that there should be no net population transfer in regions of zero NAC. Further these two definitions can be used either in the full CTMQC scheme where there are XF corrections in both electronic and nuclear equations of motion, or in a surface-hopping scheme CTSH where they appear only in the electronic equation. The quantum momentum may also be computed via auxiliary trajectories rather than coupled trajectories, which is done in the surface-hopping scheme SHXF.

The effects of these different methods were studied here on two cases of dynamics through Tully's model of an extended coupling region between two BO surfaces. In one case, the incoming momentum of the nuclear wavepacket is high enough that there is  a single pass through the NAC region with significant transfer of population to the other BO surface, and we showed that while CTMQC with the modified definition yields accurate population transfer and coherence properties, CTMQC with the original definition of the quantum momentum suffers from the violation of the condition. The original definition however yields accurate population transfer and coherences when used within surface-hopping framework, provided the fraction of trajectories measure is used, but at the expense of violating internal consistency. This is because the hopping probability is only non-zero in regions of the NAC, so in this sense, takes care of the condition. This is true both for when the quantum momentum is computed using coupled trajectories, or from auxiliary trajectories. 

In the second case, the incoming momentum is low enough that part of the wavepacket that transfers to the upper surface, reflects and recrosses the NAC region a second time. Here CTMQC with the modified definition does a good job, although underestimates the second rise in coherence, while the other methods are less accurate and also less  in agreement with each other. Using the modified definition in a surface-hopping approach however yields poor electronic populations likely due to the incorrect forces on the nuclei from the BO surfaces in the region of interaction, and incorrect imposition of energy conservation on an individual trajectory level, that lead the trajectories to spend longer in these regions. 
Using the original definition of the quantum momentum leads to unphysical population transfer.  Again the fraction of trajectories measure reduces the error and, temporarily, gives a reasonable trend.

The method that is most well-grounded from a theoretical point of view is CTMQC, and, with the modified definition of the quantum momentum, gives reliable results for both the populations and coherences. From a practical standpoint, it has the advantage of needing fewer trajectories for convergence, and no adjustable parameters. However, the method can be more numerically challenging to deal with than methods like surface hopping.
In CTSH the forces that propagate the trajectories are purely adiabatic, and the NAC vectors, which are computational expensive for molecular systems, are not needed along the dynamics unless the velocity rescaling after a surface hop has occurred is performed along the direction of the corresponding NAC vector.   
But with coupled trajectories, patience is required to wait for all the trajectories at each time-step in order to compute the coupled-trajectory term.
The computational advantage of SHXF is evident due to its use of independent trajectories supplemented with auxiliary trajectories. 
So, in addition to the first-principles nature of the electronic equation, SHXF has an attraction due to its independent trajectories, but surface-hopping in itself is unsatisfactory due to the {\it ad hoc} aspects such as the very hopping procedure, choice of velocity adjustment, energy conservation on the independent trajectory level, and the choice of how to deal with forbidden hops. These problems contribute to the internal consistency, which is generally partly reduced by the XF-based quantum momentum term as seen in previous studies on decoherence, but we have seen here that it can also exacerbate it. Still, even then, we can sometimes {\it temporarily} take advantage of this internal inconsistency by using the fraction of trajectories measure, since it takes care of the condition that there should be zero net population transfer when there is zero NAC; this is only temporary since the evolving populations will continue to be wrong, and will give poor results when the system then meets another interaction region. 

In conclusion, the XF offers a new playground for the development of non-adiabatic dynamics of electrons and ions, and we hope this work will be helpful to inform future adventures in this area.

\bibliographystyle{ieeetr}
\bibliography{./ref_arxiv.bib,./others_arxiv.bib}

\end{document}